\documentclass[pre,twocolumn,amsmath,amssymb]{revtex4}
\usepackage{graphicx}% Include figure files
\usepackage{dcolumn}% Align table columns on decimal point
\usepackage{bm}% bold math

\begin{document}

\newcommand{\bfomega}{\mbox{\boldmath $\bf \omega$}}
\newcommand{\bfGamma}{\mbox{\boldmath $\bf \Gamma$}}
\newcommand{\cald}{\tilde{d}}
\newcommand{\eqspaA}{\hspace{0.5cm}}
\newcommand{\eqspaB}{\hspace{0.5cm}}

\newcommand{\sect}[1]
   {\vspace{0.3cm} $\langle$ {\it #1} $\rangle$ }  
\makeatletter 
   \def\@biblabel#1{[#1]} 
   \def\refname{\vspace{0cm} 
   \noindent {\it References}  
   \vspace{0cm}} 
   \makeatother    
\renewcommand{\thefootnote}{\fnsymbol{footnote}}

\newcommand{\vspfigA}{\vspace{0.25cm}}  
\newcommand{\vspfigB}{\vspace{0.25cm}}  
\newcommand{\widthfigA}{0.4\textwidth}  
\newcommand{\widthfigB}{0.38\textwidth} 

\baselineskip 0.395cm

%%%%%%%%%%%%%%%%%%%%%%%%%%%%%%%%%%%%%%%%%%%%%%%%%%%%%%%%%%%%%%%%%%%%%%

\title{Steady shear flow thermodynamics based on 
a canonical distribution approach
 }

\author{Tooru Taniguchi and Gary P. Morriss}

\affiliation{School of Physics, University of New South Wales, 
             Sydney, New South Wales 2052, Australia} 
 
\date{\today}

\begin{abstract}

   A non-equilibrium steady state thermodynamics 
to describe shear flows is developed 
using a canonical distribution approach. 
   We construct a canonical distribution for shear flow 
based on the energy in the moving frame using 
the Lagrangian formalism of the classical mechanics. 
   From this distribution we derive the Evans-Hanley 
shear flow thermodynamics, which is characterized by 
the first law of thermodynamics $d\mathcal{E} = T d\mathcal{S} - 
\mathcal{Q}d\gamma$ relating infinitesimal changes in 
energy $\mathcal{E}$, entropy $\mathcal{S}$ 
and shear rate $\gamma$ with kinetic temperature $T$. 
   Our central result is that the coefficient $\mathcal{Q}$ 
is given by Helfand's moment for viscosity. 
   This approach leads to thermodynamic stability 
conditions for shear flow, one of which is equivalent to 
the positivity of the correlation 
function of $\mathcal{Q}$. 
   We emphasize the role of the external work required to sustain 
the steady shear flow in this approach, and show theoretically 
that the ensemble average of its power $\dot{W}$ 
%$-\gamma P_{xy}$ with shear stress $P_{xy}$ 
must be non-negative. 
   A non-equilibrium entropy, increasing in time, is introduced, 
so that the amount of heat based on this entropy 
is equal to the average of $\dot{W}$.
   Numerical results from non-equilibrium molecular dynamics 
simulation of two-dimensional many-particle systems with soft-core 
interactions are presented which support our interpretation.   

%   We also discuss that the pressure is a increasing 
%function of the shear rate. 

\vspace{0.5cm}
%\pacs{
\noindent
Pacs numbers: 05.70.Ln, 
% Nonequilibrium and irreversible thermodynamics 
% (see also 82.40.Bj Oscillations, chaos, and bifurcations
% in physical chemistry and chemical physics)
83.50.Ax, 
%Steady shear flows, viscometric flow
05.20.Jj, 
%Statistical mechanics of classical fluids 
%(see also 47.10. 1g General theory in fluid dynamics)
83.60.Rs 
%Shear rate-dependent structure 
%(shear thinning and shear thickening)
%---
%
%47.70.Nd Nonequilibrium gas dynamics
%
%47.20.Ft Instability of shear flows
%
%05. Statistical physics, thermodynamics, and nonlinear 
%dynamical systems (see also 02.50. 2r Probability theory, 
%stochastic processes, and statistics)
%
%83. Rheology (see also section 47 Fluid dynamics)
%---
\vspace{1.3cm} 
%\vspace{1.5cm}
\end{abstract}

\maketitle  

%%%%%%%%%%%%%%%%%%%%%%%%%%%%%%%%%%%%%%%%%%%%%%%%%%%%%%%%%%%%%%%%%%%%%%

\section{Introduction}

   The great success of thermodynamics as a physical theory 
to describe various equilibrium phenomena has stimulated attempts 
to generalize it to a theory applicable to macroscopic time-dependent phenomena, 
namely to a non-equilibrium thermodynamics. 
   Many efforts have been devoted to this subject, 
and led to some proposals for non-equilibrium thermodynamics, 
for example, the classical irreversible thermodynamics 
%based on the local equilibrium assumption 
\cite{Gro84}, the rational thermodynamics \cite{Tru84,Lav93} 
and the extended irreversible thermodynamics \cite{Jou01}. 
   Recently, a non-equilibrium thermodynamics, which tries to give 
more rigorous predictions by restricting its applied field 
into non-equilibrium steady states, is also discussed 
\cite{Ono98,Hat01,Sas01}. 

   Shear flow is a typical example of non-equilibrium steady 
phenomena. 
   For a constant velocity gradient it has a steady current (the shear 
stress), and 
has many applications in the investigation of rheological properties 
of materials \cite{Bar89,Tsh89}. 
   Such models have been widely used to calculate the shear viscosity, 
whose shear rate dependence is still actively discussed 
\cite{Tra95,Mat00,Mar01a,Ge01}. 
   The apparent existence of critical phenomenon, appearing as a transition 
from a uniform bulk phase to an organized string-like phase, 
is shown at high shear rate \cite{Erp84,Woo84,Hey85}. 
   It is used as an application of non-Hamiltonian 
molecular dynamics 
called the non-equilibrium thermostated dynamics \cite{Eva90a}, 
by which some non-equilibrium dynamical properties 
such as the conjugate pairing rule for the Lyapunov spectrum 
\cite{Dre88,Eva90b,Det96a,Tan02} 
and the fluctuation theorem \cite{Eva93,Gal95} were discussed. 
   Shear flow has also been described 
by the Bhatnagar-Gross-Krook kinetic equation, 
which is a simplification of the  Boltzmann equation, 
and using this equation, transport coefficients 
and hydrodynamic modes were calculated \cite{Zwa79,San91,Lee97}. 
   Steady shear flow is a spatially homogenous 
and time-independent phenomenon, so a simple description compared to 
general non-equilibrium phenomena may be expected.
   However, although it is such a simple steady non-equilibrium 
phenomenon, a convincing thermodynamic description 
of shear flow is not known yet. 

   A proposed non-equilibrium steady state thermodynamics 
to describe shear flow was given by Evans and Hanley 
\cite{Eva80a,Eva80b,Han82,Eva83}. 
   It expresses the first law of thermodynamics 
for the shear flow by adding 
the term $\xi d\gamma$ expressing the 
response to a shear rate $\gamma$, namely 

\begin{eqnarray}
   d\mathcal{E} = T d\mathcal{S} + \xi d\gamma 
\label{EvansHanley}\end{eqnarray}

\noindent as a relation among infinitesimal quasi-static changes of 
internal energy $\mathcal{E}$, entropy $\mathcal{S}$ 
and the shear rate $\gamma$ with temperature $T$ and 
the coefficient $\xi$ defined by 
$\xi\equiv \partial \mathcal{E} 
/ \partial \gamma |_{\mathcal{S}}$ \cite{noteI1}. 
   As a conceptual feature, the Evans-Hanley thermodynamics is 
characterized by the fact that  
the shear rate is an external parameter 
chosen as an additional variable 
to describe non-equilibrium effects. 
   This is analogous to the choice of variables 
in equilibrium thermodynamics where the variables are 
chosen as parameters manipulated externally, for example, the 
temperature and volume, etc.
  This choice of thermodynamic variables has the advantage 
that observables in 
this approach are rather easy to access by experiments and 
computer simulations. 
   This feature also distinguishes the Evans-Hanley thermodynamics 
from some other non-equilibrium thermodynamics, 
%like the classical irreversible thermodynamics,    
in which local quantities related to conserved quantities, 
such as the local momentum density,  
are chosen as additional variables to describe 
non-equilibrium effects, 
because they change slowly with time and are consistent 
with the phenomenological equations of hydrodynamics. 
   Compared with such a general formalism of non-equilibrium 
thermodynamics, the Evans-Hanley thermodynamics gives 
a much simpler description, as its applied field 
is restricted to steady shear flow. 
   On the other hand, one of the problems 
in the Evans-Hanley shear flow thermodynamics was that 
a clear physical meaning for the coefficient $\xi$, 
especially its microscopic expression, was not known, so that 
one has not had clear experimental or numerical evidence 
to support this thermodynamics. 
   On this point, Ref. \cite{Eva89} tried to calculate 
numerically a value of $\xi$ by introducing a non-equilibrium entropy 
in a low density system.
   Also recently, Ref. \cite{Dai03} discussed 
a phenomenological expression 
for $\xi$ using the Maxwell model for shear flow.

   Another important aspect of non-equilibrium thermodynamics is 
its construction from a solid statistical mechanical foundation. 
   Some attempts in this direction have been discussed using 
the non-equilibrium canonical distribution approach 
\cite{Mor56,Kub57,Mor58,Yam67,Kaw73,Zub74}
and the projection operator approach
\cite{Zwa61,Mor65,Rob65,Rob67,Gro82,Tan97} 
and so on.
%to describe non-equilibrium thermodynamical phenomena 
%in a statistical mechanical sense 
%\cite{Mor56,Kub57,Mor58,Yam67,Kaw73,Zub74,noteI2}. 
   As one such approach, 
the non-equilibrium canonical distribution approach 
justifies the response formula 
for thermodynamic perturbations, which is different from 
a mechanical perturbation expressed 
as a change in an external parameter 
appearing explicitly in the Hamiltonian \cite{Kub78}. 
   It uses, in principle, the distribution $f(\bfGamma)$ 
of canonical type 

\begin{eqnarray}
   f(\bfGamma) = \Xi^{-1} \exp\left\{ 
   -\beta\left[ H(\bfGamma) 
   + \sum_{\alpha=1}^{\tilde{n}} 
   \mu_{\alpha}(\bfGamma) A_{\alpha}(\bfGamma) 
   \right]\right\}
\label{CanonDistr0}\end{eqnarray} 

\noindent where $\Xi$ is a normalization constant, 
$\beta$ is the inverse temperature, $H(\bfGamma)$ is the Hamiltonian 
as a function of the phase space vector $\bfGamma$, and 
$\mu_{\alpha}(\bfGamma)$ and $A_{\alpha}(\bfGamma)$ are pairs of 
conjugate variables whose forms depend 
on the non-equilibrium phenomena under consideration. 
   In many cases, the distribution $f(\bfGamma)$, and therefore 
the functions $\mu_{\alpha}(\bfGamma)$ and $A_{\alpha}(\bfGamma)$, 
$\alpha=1,2,\cdots,\tilde{n}$, is introduced based on the "local 
equilibrium assumption" \cite{Zub74}, although it is not always 
necessary. 
   Using the distribution $f(\bfGamma)$ and 
the Liouville operator $\hat{L}$, we calculate average quantities 
as the ensemble average under time-evolution of the  distribution function 

\begin{eqnarray}
   \tilde{f}(\bfGamma,t) 
   &\equiv& \exp \left\{-i\hat{L} (t-t_{0})
   \right\} f(\bfGamma),  
\label{GenerShearCanon0}\end{eqnarray}

\noindent which may be regarded as the distribution function 
at time $t$ evolved from the initial canonical distribution function 
$f(\bfGamma)$ at time $t_{0}$. 
   In this way, we can derive the thermal response formula for 
viscosity, thermal conductivity and so on 
\cite{Mor58,Yam67,Kaw73,Zub74}. 
   This approach was generalized to non-Hamiltonian systems such as 
the Sllod equation for shear flow systems with isokinetic thermostat 
\cite{Mor85,Mor88,Sar98}, and 
was used to calculate some thermal quantities, 
such as the specific heat of 
non-equilibrium steady states \cite{Eva87a,Eva87b}. 
   However it is still an open problem to construct 
the Evans-Hanley shear flow thermodynamics 
from this non-equilibrium canonical distribution approach 
based on distributions of this type (\ref{CanonDistr0}). 
   Usually, the thermodynamic relation for 
the first law of thermodynamics in the canonical distribution approach 
is introduced based upon the local equilibrium assumption, 
but the form (\ref{EvansHanley}) 
as the first law of thermodynamics for shear flow 
cannot be justified by this assumption, because Eq. (\ref{EvansHanley}) 
including the term $\xi d\gamma$ expressing 
the non-equilibrium effect of the shear flow cannot be
attributed to an equilibrium thermodynamical relation 
even if we consider a very small portion of system.

   The principal aim of this paper is to derive the Evans-Hanley 
shear flow thermodynamics based on the canonical distribution approach 
to non-equilibrium steady states. 
   First we show that the canonical distribution for shear flow 
is represented by the distribution (\ref{CanonDistr0}) in 
the case where $\tilde{n}=1$, $\mu_{\alpha}(\bfGamma)=\gamma$ and 
$A_{\alpha}(\bfGamma) = -Q(\bfGamma)$. 
   Here $Q(\bfGamma)$ is the Helfand moment of viscosity and 
its time-derivative is connected with the off-diagonal component 
of the pressure tensor. 
   The canonical distribution given here is different 
from the canonical distribution based on the local equilibrium 
assumption, because the "non-equilibrium term" 
$-\gamma Q(\bfGamma)$ in the distribution $f(\bfGamma)$ 
cannot be neglected even if we take the system size 
to be very small. 
   So to derive the canonical distribution for shear flow, 
we introduce the Hamiltonian for the moving frame which follows the 
steady global current, then using the Lagrangian 
techniques of classical mechanics, and the fact that the quantity 
$H(\bfGamma) + \sum_{\alpha=1}^{\tilde{n}} \mu_{\alpha}(\bfGamma) A_{\alpha}(\bfGamma)$ 
in the distribution (\ref{CanonDistr0}) should correspond to this 
moving frame Hamiltonian. 
   This procedure, which is one of the important points of this paper, 
gives a systematic way to choose the function 
$\sum_{\alpha=1}^{\tilde{n}} \mu_{\alpha}(\bfGamma)A_{\alpha}(\bfGamma)$ 
for the canonical distribution approach to non-equilibrium steady states. 
   To justify this procedure concretely, we show that by applying it 
to the rotating system we can obtain the well known canonical distribution 
for the rotating system, and therefore its thermodynamics. 
%   It is indicated that there are some common features in 
%the rotating system and the shear flow system; 
%for example, both the systems have a simple external parameter 
%such as the angular velocity in the rotating systems and shear rate 
%in the shear flow systems. 

   Second we show that our approach 
is consistent with the linear response formula for viscosity, 
which can be derived from the ensemble average of the shear stress 
using the distribution (\ref{GenerShearCanon0}). 
   To interpret the derivation of this response formula 
we emphasize the role of the work required 
to sustain the steady shear flow. 
   We introduce a non-equilibrium entropy, which increases in time, 
and show that the heat based on this entropy has the 
same magnitude as the power needed to sustain the shear flow. 

   Third we derive the form (\ref{EvansHanley}) of the first law 
of thermodynamics for shear flow from our canonical distribution 
approach, and show that the quantity $\xi$ in the 
form (\ref{EvansHanley}) is given by $\xi=-\overline{Q}$ 
with $\overline{Q}$ being the ensemble average of the 
Helfand moment for viscosity $Q(\bfGamma)$. 
   We also discuss the thermodynamic stability condition for 
shear flow, which leads to the positivity 
of the correlation function 
of $Q(\bfGamma)$ as well as the positivity of the specific heat. 

   Finally we present some numerical calculations of many-particle 
systems with soft-core interactions to support our thermodynamic
interpretation of steady shear flow. 
   Here, we use the Sllod equations with the isokinetic thermostat 
and Lee-Edwards boundary condition \cite{Lee72}. 
   In these simulations we show the shear rate dependences of  
the average of Helfand's moment of viscosity, 
its correlation function and the work needed to sustain the shear flow.

%%%%%%%%%%%%%%%%%%%%%%%%%%%%%%%%%%%%%%%%%%%%%%%%%%%%%%%%%%%%%%%%%%%%%%

\section{Canonical Distribution for Steady Flows}
\label{CanDisSteFlo}

\subsection{Moving Frame and Energy}

   Systems discussed in this paper are steady flows, 
in which the global velocity distribution of the flow 
is given by a time-independent function 
$\mathbf{V}(\textbf{r})$ 
at the position $\textbf{r}$ in the inertial frame 
$\mathcal{F}^{(ine)}$. 
   We consider a system which consists of 
$N$ particles with the same mass $m$ 
and is described by classical mechanics 
(without magnetic fields). 

   In this section we use the Lagrangian formalism 
of classical mechanics to compare quantities in different frames.  
The Lagrangian formalism is a direct consequence of 
the frame-independent principle of least action 
$\delta \int_{t_{1}}^{t_{2}} dt L  = 0$ 
for fixed values of positions at 
times $t_{1}$ and $t_{2}$, 
where $L$ is the Lagrangian 
as a function of particle positions and velocities. 

   In the inertial frame $\mathcal{F}^{(ine)}$ 
the Lagrangian $L^{(ine)} 
= L^{(ine)}(\mathbf{v}^{(ine)},\mathbf{q})$ is given by 

\begin{eqnarray}
   L^{(ine)}(\mathbf{v}^{(ine)},\mathbf{q}) 
   = \frac{1}{2} m \left| {\mathbf{v}^{(ine)}} 
   \right|^{2} - U(\mathbf{q})
\label{LagraIne}\end{eqnarray}

\noindent as a function of $\mathbf{v}^{(ine)}\equiv 
(\mathbf{v}_{1}^{(ine)},\mathbf{v}_{2}^{(ine)},
\cdots,\mathbf{v}_{N}^{(ine)})$ and $\mathbf{q}\equiv
(\mathbf{q}_{1},\mathbf{q}_{2},\cdots,\mathbf{q}_{N})$,  
where $\mathbf{v}_{j}^{(ine)}$ and $\mathbf{q}_{j}$ are  
the velocity and the spatial position of the $j$-th particle, 
respectively,
% in the inertial frame $\mathcal{F}_{(ine)}$, 
and $U(\mathbf{q})$ is the potential function 
as a function of $\mathbf{q}$. 
   Using the definitions 
$\mathbf{p}^{(ine)} 
\equiv \partial L^{(ine)}/\partial \mathbf{v}^{(ine)}$ 
and $H^{(ine)}(\mathbf{p}^{(ine)},\mathbf{q}) 
\equiv \mathbf{p}^{(ine)} \cdot\mathbf{v}^{(ine)} - L^{(ine)}$, 
the Lagrangian (\ref{LagraIne}) 
leads to expressions for the momentum $\mathbf{p}^{(ine)}$ 
and the Hamiltonian $H^{(ine)}$ as 

\begin{eqnarray}
   \mathbf{p}^{(ine)} 
   = m \mathbf{v}_{j}^{(ine)}
\label{MomenIne}\end{eqnarray}
\begin{eqnarray}
   H^{(ine)}(\mathbf{p}^{(ine)},\mathbf{q}) 
   = \frac{1}{2m}  \left| \mathbf{p}^{(ine)} \right|^{2} 
     + U(\mathbf{q}) .
\label{HamilIne}\end{eqnarray}

\noindent Here we note that the Hamiltonian $H^{(ine)}$ 
is a function of $\mathbf{p}^{(ine)}$ and $\mathbf{q}$, 
namely $H^{(ine)} 
= H^{(ine)}(\mathbf{p}^{(ine)},\mathbf{q})$.

   Now, we introduce the velocity $\mathbf{v}_{j}^{(mov)}$ 
of the $j$-th particle in the moving frame $\mathcal{F}^{(mov)}$, 
which is connected to the velocity 
$\mathbf{v}_{j}^{(ine)}$ 
in the inertial frame $\mathcal{F}^{(ine)}$ by 

\begin{eqnarray}
   \mathbf{v}_{j}^{(mov)} 
   \equiv \mathbf{v}_{j}^{(ine)} - \mathbf{V}(\textbf{q}_{j}).
\label{VeloMov}\end{eqnarray}

\noindent The quantity $\mathbf{v}_{j}^{(mov)}$ is often referred to
as the thermal velocity of particle $j$. The position $\textbf{q}$ 
is invariant under this frame change  $\mathcal{F}^{(ine)}
\rightarrow \mathcal{F}^{(mov)}$. 
   Inserting the velocity 
$\mathbf{v}_{j}^{(ine)} 
=\mathbf{v}_{j}^{(mov)} + \mathbf{V}(\textbf{q}_{j})$ 
into Eq. (\ref{LagraIne}), 
the Lagrangian $L^{(mov)} 
= L^{(mov)}(\mathbf{v}^{(mov)},\mathbf{q})$ of the system 
in the moving frame  $\mathcal{F}^{(mov)}$ is given by 

\begin{eqnarray}
   &&\hspace{-0.5cm} 
   L^{(mov)}(\mathbf{v}^{(mov)},\mathbf{q}) = L^{(ine)}
   \nonumber \\
%   &&\hspace{0.5cm} 
   && =\frac{1}{2} m \sum_{j=1}^{N}\left| 
   {\mathbf{v}_{j}^{(mov)}} 
   + \mathbf{V}(\textbf{q}_{j})
   \right|^{2} - U(\mathbf{q})
\label{LagraMov}\end{eqnarray}

\noindent as a function of $\mathbf{v}^{(mov)}\equiv 
(\mathbf{v}_{1}^{(mov)},\mathbf{v}_{2}^{(mov)}
\cdots,\mathbf{v}_{N}^{(mov)})$ and $\mathbf{q}$. 
   Equation (\ref{LagraMov}) leads to the momentum 
$\mathbf{p}^{(mov)}
\equiv \partial L^{(mov)}/\partial \mathbf{v}^{(mov)}$ as 

\begin{eqnarray}
   \mathbf{p}_{j}^{(mov)} 
   &=& m \left[ \mathbf{v}_{j}^{(mov)} 
   + \mathbf{V}(\textbf{q}_{j}) 
   \right] \nonumber \\
   &=& m \mathbf{v}_{j}^{(ine)}
   = \mathbf{p}_{j}^{(ine)}. 
\label{MomenMov}\end{eqnarray}

\noindent Therefore the momentum is 
independent of choice of the frames $\mathcal{F}^{(mov)}$ 
and $\mathcal{F}^{(ine)}$, and hereafter we use 
the notation $\mathbf{p}\equiv (\mathbf{p}_{1}, 
\mathbf{p}_{2}\cdots,\mathbf{p}_{N})\equiv\mathbf{p}^{(mov)} 
= \mathbf{p}^{(ine)}$ for the momentum, and 
also use the notation 
$\bfGamma \equiv (\mathbf{p},\mathbf{q})$ 
for the phase space vector. 
   On the other hand, the Hamiltonian $H^{(mov)} 
= H^{(mov)}(\bfGamma)$ in the moving frame 
$\mathcal{F}^{(mov)}$ is given by

\begin{eqnarray}
   H^{(mov)}(\bfGamma) 
   = H^{(ine)}(\bfGamma) 
     -  \sum_{j=1}^{N}\mathbf{p}_{j} 
     \cdot \mathbf{V}(\textbf{q}_{j}).
\label{HamilMov}\end{eqnarray}

\noindent using the definition  
$H^{(mov)}(\bfGamma) 
\equiv \mathbf{p} \cdot\mathbf{v}^{(mov)} - L^{(mov)}$ 
and Eq. (\ref{HamilIne}).
   It is essential to note that although 
the momenta $\mathbf{p}^{(mov)}$ 
and $\mathbf{p}^{(ine)}$ are equal, 
the Hamiltonian $H^{(mov)}(\bfGamma)$ 
in the moving frame $\mathcal{F}^{(mov)}$ is 
different from the Hamiltonian $H^{(ine)}(\bfGamma)$ 
in the inertial frame $\mathcal{F}^{(ine)}$ and their 
difference is proportional to the global current distribution 
$\mathbf{V}(\textbf{q}_{j})$ \cite{note2}.

%%%%%%%%%%%%%%%%%%%%%%%%%%%%%%%%%%%%%%%%%%%%%%%%%%%%%%%%%%%%%%%%%%%%%%

\subsection{Canonical Distribution}

   The central assumption of this paper is that  
in the moving frame $\mathcal{F}^{(mov)}$ 
defined by the global velocity distribution 
$\mathbf{V}(\textbf{r})$ the system can be regarded 
as an equilibrium state. 
   It is important to note that this assumption is not 
obvious, because generally a global flow causes some 
local effects such as string phases in shear flows 
\cite{Erp84,Woo84,Hey85,Eva90a}
and possibly turbulent phases and so on,  
which can destroy this assumption. 
   However these phases generally occur in far-from 
equilibrium states so we may expect that our assumption 
is satisfied in a regime near equilibrium. 
   Under this assumption, we introduce the 
canonical distribution 

\begin{eqnarray}
%   &&
   f(\bfGamma) 
%   \nonumber \\
%   &&\eqspaA 
   &=& \Xi^{-1} 
   \exp\left\{-\beta H^{(mov)}(\bfGamma)\right\} 
   \label{Canon0} \\
%   &&\eqspaA 
   &=& \Xi^{-1} 
   \exp\left\{-\beta \left[
      H^{(ine)}(\bfGamma) 
      -  \sum_{j=1}^{N}\mathbf{p}_{j} 
      \cdot \mathbf{V}(\textbf{q}_{j}) 
   \right]\right\} 
   \nonumber \\
\label{Canon}\end{eqnarray}

\noindent for steady flow, where $\beta$ is the inverse 
temperature $1/(k_{B}T)$ with the Boltzmann constant $k_{B}$ 
and the temperature $T$, and $\Xi$ is the partition function 
$\Xi\equiv\int d\bfGamma 
\exp\left\{-\beta H^{(mov)}(\bfGamma)\right\}$. 
   For simplicity we use units so that $k_{B}=1$ 
hereafter in this paper. 
   It should be noted that the function $f(\bfGamma)$ 
is the distribution of $\textbf{p}^{(ine)}$ and $\textbf{q}$ 
in the inertial frame $\mathcal{F}^{(ine)}$, 
as well as the distribution of 
$\textbf{p}^{(mov)}$ and $\textbf{q}$ 
in the moving frame $\mathcal{F}^{(mov)}$, 
because of the identity of the momenta implies 
$\bfGamma = (\mathbf{p}^{(ine)},\mathbf{q}) 
= (\mathbf{p}^{(mov)},\mathbf{q})$.
 
   The canonical distribution (\ref{Canon}) is 
different from that obtained from the "local equilibrium assumption", 
which is popular 
in many texts on non-equilibrium statistical mechanics. 
%   Actually the "non-equilibrium term" 
%$-  \sum_{j=1}^{N}\mathbf{p}_{j} 
%\cdot \mathbf{V}(\textbf{q}_{j}) $ 
%in the distribution (\ref{Canon}) does not disappear, 
%even if we consider an enough small portion of the system.  
%   On the other hand, 
   In this approach the canonical 
distribution function $g(\bfGamma)$ is chosen as 

\begin{eqnarray}
   &&g(\bfGamma) 
   \nonumber \\
   && \equiv 
   \tilde{\Xi}^{-1} \exp\left\{-\beta \left[
      \sum_{j=1}^{N} \frac{1}{2m} \left| 
      \mathbf{p} - m \mathbf{V}(\textbf{q}_{j})\right|^{2}
      +U(\mathbf{q})
   \right]\right\}
   \nonumber \\
   &&= \tilde{\Xi}^{-1} 
   \exp\left\{-\beta \left[
     H^{(mov)}(\bfGamma) 
     + \sum_{j=1}^{N} \frac{1}{2} m  
      \left|\mathbf{V}(\textbf{q}_{j})\right|^{2}
   \right]\right\}
   \nonumber \\
\label{CanonLocalEquib}\end{eqnarray}
   
\noindent with $\tilde{\Xi}\equiv\int d\bfGamma 
\exp\{-\beta [\sum_{j=1}^{N} | 
\mathbf{p} - m \mathbf{V}(\textbf{q}_{j}) |^{2}/(2m)
+ U(\mathbf{q})]\}$. 
   This type of distribution was actually used 
to calculate the shear viscosity 
\cite{Zwa79,San91,Mor56,Yam67,Kaw73}, 
and its localized version was used as a 
canonical distribution under the local equilibrium assumption 
(See for example, Ref. \cite{Zub74}). 
   However it is not consistent with the thermodynamics of 
rotating systems discussed in the next subsection, so 
the difference between the two distributions 
(\ref{Canon}) and (\ref{CanonLocalEquib}) 
is crucial to the subject of this paper, that is 
a statistical foundation for steady state thermodynamics. 
   Another problem with the distribution $g(\bfGamma)$ is
that it does not take into account the inertial force. 
   Despite these problems, 
one may also notice that the deviation of the distribution 
(\ref{CanonLocalEquib}) from the distribution (\ref{Canon}) 
is of order $\mathcal{O}(|\mathbf{V}|^{2})$ 
in the global velocity, and near equilibrium 
it may give small effects compared with the effects 
given by the second term on the right-hand side of 
Eq. (\ref{HamilMov}) which is of order 
$\mathcal{O}(|\mathbf{V}|)$.

%%%%%%%%%%%%%%%%%%%%%%%%%%%%%%%%%%%%%%%%%%%%%%%%%%%%%%%%%%%%%%%%%%%%%%

\section{Rotating Systems and their Thermodynamics}
\label{ThermRotat} 

   Before discussing the canonical distribution approach 
to shear flow systems, in this section we give 
a derivation of the well known canonical distribution and 
the thermodynamics for uniformly rotating systems,  
based on the formalism given in Sec. \ref{CanDisSteFlo}. 
   The rotating system is a typical example to which our approach 
can be applied, and a comparison between the rotating system 
and the shear flow system  
gives a useful insight into the canonical distribution approach 
and the thermodynamics of shear flow systems 
which will be developed in Secs. \ref{ShearFlowCano} 
and \ref{ShearFlowsTherm}.

%%%%%%%%%%%%%%%%%%%%%%%%%%%%%%%%%%%%%%%%%%%%%%%%%%%%%%%%%%%%%%%%%%%%%%

\subsection{Canonical Distribution for Rotating Systems}

   We consider a rotating system with a constant angular 
velocity vector $\bfomega$. 
   We assume that the Hamiltonian $H^{(ine)}(\bfGamma)$
%total momentum of the system is 
%zero and the potential function  $U(\mathbf{q})$
is invariant under rotation about the axis of rotation. 
   In this section the origin of the spatial coordinates 
and the axis of rotation
is taken at the center of mass of the system. 
   Under these conditions the global velocity distribution 
function $\mathbf{V}$ is given by 

\begin{eqnarray}
   \mathbf{V}(\mathbf{q}_{j}) = \bfomega\times \mathbf{q}_{j} 
\label{RotGloVel}\end{eqnarray}

\noindent where $\times$ is the usual vector product. 
   This global velocity distribution can be sustained 
without any external effect in isolated systems, 
because of the conservation of the total angular momentum. 

   Using Eq. (\ref{RotGloVel}), relation (\ref{HamilMov}) 
%to connect the Hamiltonian $H^{(mov)}(\bfGamma)$ 
%in the moving frame $\mathcal{F}^{(mov)}$ 
%with the Hamiltonian $H^{(ine)}(\bfGamma)$ 
%in the moving frame $\mathcal{F}^{(ine)}$  
is rewritten as 

\begin{eqnarray}
   H^{(mov)}(\bfGamma) 
   = H^{(ine)}(\bfGamma) 
     -  \bfomega \cdot \mathbf{M}(\bfGamma) 
\label{RotHamilMov}\end{eqnarray}

\noindent with the total angular momentum 
%$\mathbf{M}(\bfGamma)$
%
%\begin{eqnarray}
$
   \mathbf{M}(\bfGamma)
   \equiv \sum_{j=1}^{N}\textbf{q}_{j}\times\mathbf{p}_{j}.
$
%\label{RotHamilMov}\end{eqnarray}
%
%\noindent 
In comparison with shear flow systems 
discussed in Sec. \ref{ShearFlowCano}, it is important to note 
that the total angular momentum 
$\mathbf{M}(\bfGamma)$ 
is conserved in the inertial frame 
$\mathcal{F}^{(ine)}$, namely 

\begin{eqnarray}
   i \hat{L}^{(ine)} \mathbf{M}(\bfGamma) = 0 ,
\label{RotMomenConse}\end{eqnarray}

\noindent where 
$\hat{L}^{(ine)}$ is the Liouville operator defined by 
$i\hat{L}^{(ine)} X $ $\equiv 
(\partial H^{(ine)}(\bfGamma)
   /\partial \mathbf{p})
   \cdot (\partial X /\partial \mathbf{q}) 
$ $ - $ $(\partial H^{(ine)}(\bfGamma)
   /\partial \mathbf{q})
   \cdot (\partial X /\partial \mathbf{p}) $ 
for any function $X$ of $\bfGamma$. 
%   Equation (\ref{RotMomenConse}) is a mathematical expression of 
%the conservation law of the total momentum. 
   Using Eqs. (\ref{Canon}) and (\ref{RotHamilMov}) 
the canonical distribution for the rotating system 
is represented as \cite{Lan59} 

\begin{eqnarray}
   f(\bfGamma) 
   = \Xi^{-1} 
   \exp\left\{-\beta \left[
      H^{(ine)}(\bfGamma) 
      - \bfomega \cdot \mathbf{M}(\bfGamma) 
   \right]\right\} . 
\label{RotCanon}\end{eqnarray}

\noindent The distribution (\ref{RotCanon}) is stationary, 
namely $i \hat{L}^{(ine)} f(\bfGamma) 
= i \hat{L}^{(mov)} f(\bfGamma) = 0$, 
in both frames $\mathcal{F}^{(mov)}$ 
and $\mathcal{F}^{(ine)}$. 
%because of $i \hat{L}^{(ine)} H^{(ine)}(\bfGamma) 
%= 0$ for the moving frame $\mathcal{F}^{(mov)}$, 
%and $i \hat{L}^{(ine)} H^{(ine)}(\bfGamma) 
%= 0$ and Eq. (\ref{RotMomenConse}) 
%for the inertial frame $\mathcal{F}^{(ine)}$. 
   Here $\hat{L}^{(mov)}$ is the Liouville operator 
in the moving frame $\mathcal{F}^{(mov)}$ and is 
defined by 
$i\hat{L}^{(mov)} X $ $\equiv 
(\partial H^{(mov)}(\bfGamma)
   /\partial \mathbf{p})
   \cdot (\partial X /\partial \mathbf{q}) 
$ $ - $ $(\partial H^{(mov)}(\bfGamma)
   /\partial \mathbf{q})
   \cdot (\partial X /\partial \mathbf{p}) $ 
for any function $X(\bfGamma)$, similarly to $\hat{L}^{(ine)}$. 
   The distribution (\ref{RotCanon}) has the 
general form (\ref{CanonDistr0}) 
of the canonical distribution  
in the case that   
$H(\bfGamma) = H^{(ine)}(\bfGamma)$, $\tilde{n} = \cald$, and 
$A_{j}(\bfGamma)$ is the component of $\mathbf{M}(\bfGamma)$, and 
$\mu_{j}(\bfGamma)$ is the component of $\bfomega$ 
in the $\cald$-dimensional system. 

   It is valuable to note that from the canonical distribution 
(\ref{RotCanon}) we can derive the distribution function 
$f'(\mathbf{v}^{(mov)},\mathbf{q})$
for the position $\mathbf{q}$ and the velocity 
$\mathbf{v}^{(mov)}$ in the moving frame $\mathcal{F}^{(mov)}$ as 

\begin{eqnarray}
   &&f'(\mathbf{v}^{(mov)},\mathbf{q}) 
   \nonumber \\
   && = \Xi^{-1} 
   \exp\left\{-\beta \left[
      \sum_{j=1}^{N} \frac{1}{2} m \left| 
         \mathbf{v}_{j}^{(mov)} \right|^{2} + U(\mathbf{q})
      + u(\textbf{q}) 
   \right]\right\} 
   \nonumber \\
\label{RotCanon2}\end{eqnarray}

\noindent using Eq. (\ref{MomenMov}), where 
$u(\textbf{q})$ is given by 
$u(\textbf{q}) \equiv - \sum_{j=1}^{N} \frac{1}{2} m  
\left|\bfomega\right|^{2} r_{j}^{2}$
with $r_{j} \equiv |\bfomega
\times \mathbf{q}_{j}|/|\bfomega|$. 
   The function $f'(\mathbf{v}^{(mov)},\mathbf{q})$ is 
the distribution for a rotating system  
including explicitly the effect of the centrifugal potential 
$u(\textbf{q})$. 
   We cannot derive the distribution (\ref{RotCanon2}) from 
the distribution (\ref{CanonLocalEquib}).
% under the assumption of the local equilibrium. 

%%%%%%%%%%%%%%%%%%%%%%%%%%%%%%%%%%%%%%%%%%%%%%%%%%%%%%%%%%%%%%%%%%%%%%

\subsection{Thermodynamics for Rotating Systems}
\label{ThermoRotat}

   In this paper we use the notation $\overline{X}$ for the ensemble 
average of any function $X(\bfGamma)$ using 
the canonical distribution $f(\bfGamma)$, namely 

\begin{eqnarray}
   \overline{X} \equiv\int d\bfGamma X(\bfGamma) f(\bfGamma). 
\label{Avera}\end{eqnarray}

\noindent 
%   From the canonical distribution (\ref{RotCanon}) of 
%the rotating system, we define the energies $E^{(mov)}$ and 
%$E^{(ine)}$, and the averaged total momentum $\overline{\mathbf{M}}$ by    
%
%\begin{eqnarray}
%   E^{(mov)} \equiv 
%   \left\langle \; H^{(mov)}(\bfGamma)  \; \right\rangle
%   \label{RotEnergMov} 
%\end{eqnarray} 
%\begin{eqnarray}
%   E^{(ine)} \equiv
%   \left\langle \; H^{(ine)}(\bfGamma)  \; \right\rangle
%   \label{RotEnergIne}
%\end{eqnarray} 
%\begin{eqnarray}
%   \overline{\mathbf{M}} \equiv
%   \left\langle \; M(\bfGamma) \; \right\rangle
%   \label{RotMomenAve} .
%\end{eqnarray} 
%
%\noindent  where we used the notation 
%$\langle X(\bfGamma)\rangle$ as the ensemble average 
%by the canonical distribution (\ref{RotCanon}) 
%for any function $X(\bfGamma)$ of $\bfGamma$, namely 
%$\langle X(\bfGamma)\rangle\equiv\int d\bfGamma 
%X(\bfGamma) f(\bfGamma)$. 
   Using this notation and Eq. (\ref{RotHamilMov}) we obtain  
the relation 

\begin{eqnarray}
   \overline{H^{(mov)}} = \overline{H^{(ine)}} - 
   \bfomega\cdot\overline{\mathbf{M}} ,
\label{RotRelatEnerg}\end{eqnarray}

\noindent which connects the average 
energies $\overline{H^{(mov)}}$ 
and $\overline{H^{(ine)}}$ in the two different frames 
$\mathcal{F}^{(mov)}$ and $\mathcal{F}^{(ine)}$, respectively.  
%using the angular velocity and the total momentum.  

   Now we introduce the observable $S(\bfGamma)$ 
corresponding to entropy as 

\begin{eqnarray}
   S(\bfGamma) \equiv - \ln \left\{f(\bfGamma) \right\} ,
\label{ObserEntro}\end{eqnarray}
   
\noindent so that the entropy $\overline{S}$ is given by 

\begin{eqnarray}
   \overline{S} &=& \ln\Xi + \beta \overline{H^{(mov)}} 
   \label{RotEntropy2} \\
   &=& \ln\Xi + \beta \left[ 
      \overline{H^{(ine)}} 
      - \bfomega\cdot\overline{\mathbf{M}} \right]
\label{RotEntropy3}\end{eqnarray} 

\noindent  where we used Eq. (\ref{RotCanon}). 
   The free energy $F^{(mov)}$ in the moving frame 
$\mathcal{F}^{(mov)}$ 
is also introduced as  

\begin{eqnarray}
   F^{(mov)}
   &\equiv& \overline{H^{(mov)}} - T \overline{S}
   \label{RotFreEneMov0} \\
   &=& - T\ln\Xi
\label{RotFreEneMov1}\end{eqnarray} 

\noindent where we used Eq. (\ref{RotEntropy2}) to derive 
%the second equations in 
Eq. (\ref{RotFreEneMov1}).  
   Similarly the free energy $F^{(ine)}$ in the inertial 
frame $\mathcal{F}^{(ine)}$ is also introduced as 

\begin{eqnarray} 
   F^{(ine)} &\equiv& \overline{H^{(ine)}} - T \overline{S} 
   \label{RotFreEneIne0} \\
   &=& F^{(mov)}+\bfomega\cdot\overline{\mathbf{M}}
   \label{RotFreEneIne1} \\ 
%= - T \ln\Xi + \bfomega \cdot d\overline{\mathbf{M}} 
   &=& - T \left( \ln\Xi - \bfomega\cdot
       \frac{\partial\ln\Xi}{\partial \bfomega} \right)
\label{RotFreEneIne}\end{eqnarray} 

\noindent 
using Eq. (\ref{RotEntropy3}). 
   Therefore the free energies $F^{(mov)}$ and $F^{(ine)}$ 
can be calculated directly from the partition function $\Xi$.

   Noting the definition of the partition function 
$\Xi\equiv\int d\bfGamma $ $exp\{-\beta H^{(mov)}\} 
$ $= \int d\bfGamma $ 
$exp\{-\beta $ $[H^{(ine)}-\bfomega\cdot\mathbf{M}]\}$ 
including the two parameters $T (=\beta^{-1})$ and $\omega$ 
explicitly,
Eq. (\ref{RotFreEneMov1})  implies that the free energy $F^{(mov)}$ 
is a function of the temperature $T$ 
and the angular velocity $\omega$: $F^{(mov)}=F^{(mov)}(T,\omega)$, 
and we have 
   
\begin{eqnarray}
   \frac{\partial \left[\beta F^{(mov)} \right]}{\partial \beta} 
   = - \frac{\partial \ln\Xi}{\partial \beta} 
   = \overline{H^{(mov)}}
\label{RotDerFreEneMovTem}\end{eqnarray}
\begin{eqnarray}
   \frac{\partial \left[\beta F^{(mov)} \right]}{\partial \bfomega} 
   = - \frac{\partial \ln\Xi}{\partial \bfomega} 
   = - \beta\overline{\mathbf{M}}
\label{RotDerFreEneMovAng}\end{eqnarray}

\noindent Eqs. (\ref{RotDerFreEneMovTem}) and 
(\ref{RotDerFreEneMovAng}) are summarized as 

\begin{eqnarray}
   d \left[\beta F^{(mov)} \right]
   = \overline{H^{(mov)}} d\beta 
   - \beta \overline{\mathbf{M}} \cdot d\bfomega , 
\label{RotDerFreEneMov}\end{eqnarray}

\noindent where we have used the notation that $dX$ is the 
infinitesimal change in the quantity $X$.
   Inserting Eqs. (\ref{RotFreEneMov0}) and 
$d\beta = - T^{-2} dT$ 
into Eq. (\ref{RotDerFreEneMov}) we obtain 

\begin{eqnarray}
   d F^{(mov)} 
   = - \overline{S} dT 
   - \overline{\mathbf{M}} \cdot d\bfomega .
\label{RotDerFreEneMov1}\end{eqnarray}

\noindent Noting Eq. (\ref{RotFreEneIne1}), 
we also obtain 

\begin{eqnarray}
d F^{(ine)} = - \overline{S} dT 
+ \bfomega \cdot d\overline{\mathbf{M}}. 
\label{RotDerFreEneIne1}\end{eqnarray}

\noindent 
   Equation (\ref{RotDerFreEneMov1}) is also equivalent to 
   
\begin{eqnarray}
   d \overline{H^{(mov)}}
   = T d\overline{S} 
   - \overline{\mathbf{M}} \cdot d\bfomega ,
\label{RotFirstLaw1}\end{eqnarray}
\begin{eqnarray}
   d \overline{H^{(ine)}}
   = T d\overline{S} 
     +\bfomega \cdot d\overline{\mathbf{M}} ,
\label{RotFirstLaw2}\end{eqnarray}

\noindent noting the relations (\ref{RotRelatEnerg})  and 
(\ref{RotFreEneMov0}). 
   The second term on the right-hand side of 
Eq. (\ref{RotFirstLaw1}) can also be derived from 
the relation $\partial H^{(mov)}(\bfGamma)/\partial\bfomega 
= - \mathbf{M}(\bfGamma)$ derived from Eq. (\ref{RotHamilMov}), 
therefore $\partial \overline{H^{(mov)}} 
/ \partial \bfomega |_{\overline{S}} = - \overline{\mathbf{M}} $
under an adiabatic process. 
   From Eq. (\ref{RotFirstLaw1}), 
the energy $\overline{H^{(mov)}}$ in 
the moving frame $\mathcal{F}^{(mov)}$ is regarded as a 
function of $\overline{S}$ and $\bfomega$, while 
the energy $\overline{H^{(ine)}}$ in the 
inertial frame  $\mathcal{F}^{(ine)}$ is a 
function of $\overline{S}$ and $\overline{\mathbf{M}}$ 
by Eq. (\ref{RotFirstLaw2}).
   The relation (\ref{RotFirstLaw2})  
is the first law of thermodynamics for the rotating 
system, which is well known \cite{Lan59}. 

   Using Eq. (\ref{RotFirstLaw2})  
we obtain $\partial \overline{S}/\partial 
\overline{H^{(ine)}}|_{\overline{\mathbf{M}}} 
= 1/T$ and $\partial \overline{S}/
\partial\overline{\mathbf{M}}|_{\overline{H^{(ine)}}} = \bfomega/T$ 
by regarding 
the entropy $\overline{S}$ as a function of 
$\overline{H^{(ine)}}$ and $\overline{\mathbf{M}}$. 
   We notice that the thermodynamic variable conjugate 
to the averaged angular momentum 
$\overline{\mathbf{M}}$ is the inverse temperature 
times the angular velocity $\bfomega/T$, 
like the fact that the thermodynamic variable 
conjugate to the energy 
$\overline{H^{(ine)}}$ 
is the inverse temperature $1/T$. 

   After all, thermodynamic functions such as 
the free energy $F^{(mov)}$ are 
calculated from the partition function $\Xi$, and by combining 
them with the first law of thermodynamics 
we can calculate thermodynamic quantities, for example, 
the total momentum 
$\overline{\mathbf{M}} = 
-\partial F^{(mov)}/\partial\bfomega|_{T}$ 
and the entropy $\overline{S} 
= -\partial F^{(mov)}/\partial T|_{\bfomega}$, 
and their relations including the equation of state, are as 
in equilibrium thermodynamics.

%%%%%%%%%%%%%%%%%%%%%%%%%%%%%%%%%%%%%%%%%%%%%%%%%%%%%%%%%%%%%%%%%%%%%%

\section{Canonical Distribution Approach to Shear Flows}
\label{ShearFlowCano}

   In Sec. \ref{ThermRotat} we discussed the 
construction of the well-known thermodynamics of rotating systems 
from the canonical distribution (\ref{Canon}). 
   Clearly we can carry out a similar procedure for shear flow systems, 
as will be shown in Sec. \ref{ShearFlowsTherm}. 
   However it is not clear that the thermodynamic 
relations obtained by such a procedure correspond to the 
shear flow thermodynamics proposed by Evans and Hanley. 
   This section is devoted to discussing this point. 
   We compare the shear flow system with the rotating system 
discussed in Sec. \ref{ThermRotat}, and emphasize 
not only similarities but also differences between these two systems. 
   This difference leads to the necessity to consider the work 
required to sustain steady currents, and plays an essential role in 
deriving the linear response formula for viscosity in the shear flow system.

%---------------------------------------------------------------------
\subsection{Shear Flows and Helfand's Moment of Viscosity} 

   We consider the shear flow system in which the global current 
distribution is given by

\begin{eqnarray}
   \mathbf{V}(\mathbf{q}_{j}) = \gamma  q_{jy} \mathbf{i}_{x}.
\label{SheGloVel}\end{eqnarray}

\noindent where $q_{jy}$ is the $y$-component of 
the spatial coordinate $\mathbf{q}_{j}$ of the $j$-th particle, and 
$\mathbf{i}_{x}$ is the normalized vector pointing to 
the positive $x$-direction. 
   Here, $\gamma$ is the shear rate, 
namely the constant gradient of the $x$ component of the local 
velocity as a function of $y$, and is assumed to be a 
position-independent constant.

   The shear flow system was proposed to describe fluid 
filled between the two plates which move at different 
speeds, and it is frequently used to calculate viscosity. 
   The viscosity is calculated as the linear coefficient 
of the average of the $xy$-element $P_{xy}(\bfGamma)$ 
of the pressure tensor 
$P_{\alpha\beta}(\bfGamma)$ defined by 

\begin{eqnarray}
   P_{\alpha\beta}(\bfGamma) \equiv \frac{1}{\mathcal{V}} 
   \sum_{j=1}^{N} \left[
      \frac{1}{m} p_{j\alpha}p_{j\beta} 
      - q_{j\beta} 
      \frac{\partial U(\mathbf{q})}{\partial q_{j\alpha}}
   \right] 
\label{StresTenso0}\end{eqnarray}

\noindent as a function of the shear rate $\gamma$. 
   Here $\mathcal{V}$ is the volume of the system, 
$p_{j\alpha}$ ($q_{j\alpha}$) is the $\alpha$-th 
component of the momentum $\mathbf{p}_{j}$ 
(the spatial coordinate $\mathbf{q}_{j}$) of the $j$-th particle. 
   If particle-particle interactions in the system are given 
by a two-body interaction only, 
namely the potential $U(\mathbf{q})$ 
is expressed in the form $U(\mathbf{q}) = (1/2)
\sum_{j\neq k}\phi(|\mathbf{q}_{j}-\mathbf{q}_{k}|)$, 
then Eq. (\ref{StresTenso0}) can be rewritten as 

\begin{eqnarray}
   P_{\alpha\beta}(\bfGamma) 
   = \frac{1}{\mathcal{V}} 
   \sum_{j=1}^{N} 
   \left[
      \frac{1}{m} p_{j\alpha}p_{j\beta} 
%   \right.
%   \nonumber \\
%   && \left. \eqspaA 
      + \frac{1}{2} \sum_{k=1}^{N} 
      \left(q_{j\beta} - q_{k\beta} \right)  
      \Upsilon_{jk\alpha}
   \right] 
   \nonumber \\
\label{StresTenso}\end{eqnarray}

\noindent with $\Upsilon_{jk\alpha} \equiv - (1/2) 
\partial \phi(|\mathbf{q}_{j}-\mathbf{q}_{k}|) 
/ \partial q_{j\alpha}$ interpreted as 
the $\alpha$-th component 
of the force acting to the $j$-th particle 
by the $k$-th particle. 
   The pressure tensor $P_{\alpha\beta}(\bfGamma)$ 
comes from the balance equation for the momentum \cite{Eva90a}.
%   It must be noted that different from the rotating system 
%an external effect (such as moving plates holding gas or liquid) 
%is required to sustain the shear flow. 

   For the case of the global velocity 
distribution (\ref{SheGloVel}), 
Eq. (\ref{HamilMov}) is given by 

\begin{eqnarray}
   H^{(mov)}(\bfGamma) 
   = H^{(ine)}(\bfGamma) - \gamma Q(\bfGamma) .
\label{ShearHamilMov}\end{eqnarray}

\noindent where $Q(\bfGamma)$ is defined by 

\begin{eqnarray}
   Q(\bfGamma) \equiv \sum_{j=1}^{N} q_{jy} p_{jx}.
\label{Helfa}\end{eqnarray}

\noindent It is essential to note that the quantity 
$Q(\bfGamma)$ is connected to the shear stress 
$P_{xy}(\bfGamma)$ as 

\begin{eqnarray}
   i \hat{L}^{(ine)} Q(\bfGamma) = \mathcal{V} P_{xy}(\bfGamma) , 
\label{HelfaTimDer}\end{eqnarray}

\noindent namely the quantity $Q(\bfGamma)$ 
is Helfand's moment of viscosity 
\cite{Hel60,Gas98}. 
   (In the references, the name "Helfand's moment of viscosity" 
is used for the quantity $(T \mathcal{V})^{-1/2} 
Q(\bfGamma)$, but in this paper, 
for convenience we use this name for the quantity 
$Q$ itself without the 
factor $(T \mathcal{V})^{-1/2}$.) 
   Helfand's moment of viscosity is used to calculate 
the viscosity by analogy with the Einstein formula 
for the diffusion constant \cite{Ald70,Vis03}.  
   Different from the total angular momentum $\mathbf{M}(\bfGamma)$ 
appearing in Eq. (\ref{RotHamilMov}) for the rotating system, 
Helfand's moment $Q(\bfGamma)$ of viscosity 
appearing in Eq. (\ref{ShearHamilMov}) 
is not a conserved quantity, 
and this requires a different treatment of the 
canonical distribution for shear flow, as will be  
discussed in Sec. \ref{CanonDistrShear}. 

%---------------------------------------------------------------------
\subsection{Canonical Distribution for Shear Flows}
\label{CanonDistrShear}

   In the case of Eq. (\ref{SheGloVel}) the distribution function 
$f(\bfGamma)$ is given by 

\begin{eqnarray}
   f(\bfGamma) 
   = \Xi^{-1} 
   \exp\left\{-\beta \left[
      H^{(ine)}(\bfGamma) - \gamma Q(\bfGamma) 
   \right]\right\} .
\label{ShearCanon}\end{eqnarray}

\noindent This is the canonical distribution function for 
shear flow in a non-equilibrium steady state. 
   This distribution can be attributed to the 
general form (\ref{CanonDistr0}) of the canonical distribution 
in the case of  
$H(\bfGamma) = H^{(ine)}(\bfGamma)$, $\tilde{n} = 1$, 
$A_{1}(\bfGamma) = Q(\bfGamma)$ and 
$\mu_{1}(\bfGamma) = \gamma$. 

   Now we mention some physical meanings for 
the shear flow canonical distribution function (\ref{ShearCanon}). 
   For this purpose we convert the distribution function for 
(\ref{ShearCanon}) the canonical variable $\bfGamma$ 
into the distribution function 
$f''(\mathbf{v}^{(mov)},\mathbf{q})$
for the position $\mathbf{q}$ and the velocity 
$\mathbf{v}^{(mov)}$ in the moving frame $\mathcal{F}^{(mov)}$, 
and obtain 

\begin{eqnarray}
   &&f''(\mathbf{v}^{(mov)},\mathbf{q}) 
   \nonumber \\
   && = \Xi^{-1} 
   \exp\left\{-\beta \left[
      \sum_{j=1}^{N} \frac{1}{2} m \left| 
         \mathbf{v}_{j}^{(mov)} \right|^{2} + U(\mathbf{q})
      + u'(\textbf{q}) 
   \right]\right\} 
   \nonumber \\
\label{ShearCanon2}\end{eqnarray}

\noindent where the function $u'(\textbf{q})$ is defined by 

\begin{eqnarray}
   u'(\textbf{q}) 
   = - \frac{1}{2} m \gamma^{2} \sum_{j=1}^{N} q_{jy}^{2} . 
\label{ShearInePoten}\end{eqnarray} 

\noindent Here $u'(\textbf{q})$ can be regarded as a potential 
corresponding to the inertial force which pushes particles 
in the direction of the large $|q_{jy}|$ region, namely 
in the direction of the high speed region of 
the global current $\mathbf{V}$. 
   In other words, this potential $u'(\textbf{q})$ 
expresses the effect of Bernoulli's theorem 
in the hydrodynamics. 
   As another important point, using the distribution 
$f''(\mathbf{v}^{(mov)},\mathbf{q})$ we obtain 

\begin{eqnarray}
   \overline{\sum_{j=1}^{N}\frac{1}{2}m \left| 
       \mathbf{v}_{j}^{(mov)} \right|^{2}} = \frac{\cald NT}{2}
\label{KinTempe}\end{eqnarray}

\noindent for any potential $U(\mathbf{q})$, 
where $\cald$ is the spatial dimension of the system. 
%   (Note that in this paper we are using the unit 
%for the Boltzmann constant to be zero, and $\beta=T^{-1}$.) 
   Therefore $T$ can be interpreted as the kinetic temperature 
\cite{note1}. 
   It may be noted that the same relation with 
Eq. (\ref{KinTempe}) is also derived from the distribution 
$g(\bfGamma)$ defined by Eq. (\ref{CanonLocalEquib}) 
%using the local equilibrium assumption 
by interpreting the average 
$\overline{X}$ of $X$ as the average under the distribution 
$g(\bfGamma)$.

   The canonical distribution (\ref{ShearCanon}) for shear 
flow corresponds to the canonical distribution 
(\ref{RotCanon}) for the rotating system, but there is a 
significant difference between the two.  
   Owing to Eq. (\ref{HelfaTimDer}),
the distribution (\ref{ShearCanon}) is not stationary 
in the inertial frame $\mathcal{F}^{(ine)}$, namely 
$i \hat{L}^{(ine)} f(\bfGamma) \neq 0$, whereas 
the distribution (\ref{RotCanon}) is stationary 
in time in this frame. 
   (However, note that  both the distributions (\ref{RotCanon}) 
and (\ref{ShearCanon}) are stationary in time in the 
moving frame $\mathcal{F}^{(mov)}$, namely 
$i \hat{L}^{(mov)} f(\bfGamma) = 0$.)
%\cite{note2}.)
   Physically speaking, this difference comes from the fact 
that we need some work to sustain the steady current in 
the shear flow, 
whereas such work is not necessary 
in the rotational system 
because of the total angular momentum 
conservation law  
(\ref{RotMomenConse}). 
   However the effect of the work to sustain the shear flow 
is not included in the distribution (\ref{ShearCanon}) 
for the shear flow system. 
   In order to include the effect of this work 
we have to generalize the distribution (\ref{ShearCanon}), and 
introduce the distribution $\tilde{f}(\bfGamma,t)$ at time $t$ as 
   
\begin{eqnarray}
   \tilde{f}(\bfGamma,t) 
   &\equiv& \exp \left\{-i\hat{L}^{(ine)} (t-t_{0})
   \right\} f(\bfGamma) 
   \label{GenerShearCanon1} \\
   &=& f(\bfGamma) \exp\left\{ -\beta
      \gamma \mathcal{V} 
      \int_{t_{0}}^{t} ds \tilde{P}_{xy}(\bfGamma,-s+2t_{0})
   \right\} 
   \nonumber \\
\label{GenerShearCanon2}\end{eqnarray}

\noindent where $t_{0}$ is the initial time. 
%(It is taken as $t_{0}=0$ or sometimes 
%as $t_{0}\rightarrow-\infty$). 
   Here, to derive Eq. (\ref{GenerShearCanon2})  
we used the relation (\ref{HelfaTimDer}), 
$i \hat{L}^{(ine)} H^{(ine)}(\bfGamma) = 0$  
and $\exp\{-i\hat{L}^{(ine)} (t-t_{0})\} Q(\bfGamma) 
= Q(\bfGamma) - \mathcal{V} 
\int_{t_{0}}^{t} ds \tilde{P}_{xy}(\bfGamma,-s+2t_{0})$, 
and defined $\tilde{P}_{xy}(\bfGamma,t)$ 
by 

\begin{eqnarray}
   \tilde{P}_{xy}(\bfGamma,t) 
\equiv\exp\left\{i\hat{L}^{(ine)} (t-t_{0})\right\} P_{xy}(\bfGamma).
\end{eqnarray}

\noindent  
%from Eqs. (\ref{GenerShearCanon1}) and (\ref{ShearAvera}). 
   The distribution (\ref{GenerShearCanon1}) corresponds 
to the distribution (\ref{GenerShearCanon0}) in a 
general formulation of 
the non-equilibrium canonical distribution approach.
   It may be noted that the distribution $\tilde{f}(\bfGamma,t)$ 
is normalized, namely $\int d\bfGamma \tilde{f}(\bfGamma,t) = 1$,  
as well as $\int d\bfGamma f(\bfGamma) = 1$.
   Using the distribution $\tilde{f}(\bfGamma,t)$ 
we define the average 
$\langle X(\bfGamma) \rangle_{t}$ by 

\begin{eqnarray}
   \langle X(\bfGamma) \rangle_{t} 
   \equiv 
   \int d\bfGamma X(\bfGamma) \tilde{f}(\bfGamma,t) 
\label{ShearAvera}\end{eqnarray}

\noindent for any function $X(\bfGamma)$ of $\bfGamma$.
   In the rotating system it is easy to check the relation
$\tilde{f}(\bfGamma,t) = f(\bfGamma)$ and 
$\langle X(\bfGamma) \rangle_{t} = \overline{X(\bfGamma)}$ for any 
function $X(\bfGamma)$. 

   Noting Eq. (\ref{GenerShearCanon2}), the difference 
of the distribution $\tilde{f}(\bfGamma,t)$ from 
the distribution $f(\bfGamma)$ appears as the 
factor $\exp\{-\beta\gamma \mathcal{V} 
\int_{t_{0}}^{t} ds \tilde{P}_{xy}(\bfGamma,-s+2t_{0})
\}$, and we will discuss the relation of this factor to 
the work needed to sustain the shear flow  
in Sec. \ref{ShearWork}. 
   One may interpret the canonical distribution 
$f(\bfGamma)$ as a steady distribution function 
in the moving frame 
$\mathcal{F}^{(mov)}$, but in order to know about  
the work to sustain the steady flow we have to investigate it 
from the different frame $\mathcal{F}^{(ine)}$, 
%namely to investigate the inertial frame dynamics 
%$\exp \{-i\hat{L}^{(ine)} (t-t_{0}) \} f(\bfGamma)$ of 
%the moving frame steady distribution $f(\bfGamma)$, 
because the work to sustain the steady flow is 
information given by looking at the moving system 
from the inertial frame. 
   Therefore, the canonical distribution 
$f(\bfGamma)$ should not be regarded 
as an artificial test initial distribution, 
like in other  
canonical distribution approaches for a 
linear response theory \cite{Yam67,Mor58}. 
   The information about the work to sustain steady flows 
is essential to calculate 
transport coefficients such as the viscosity, as will be 
shown in Sec. \ref{GreKubForVis}.

%---------------------------------------------------------------------
\subsection{Linear Response Formula for Viscosity}
\label{GreKubForVis}

   To calculate the transport coefficient 
%such as the viscosity 
from the non-equilibrium canonical distribution approach is 
beyond the purpose of this paper. 
   However many works 
have been devoted for this subject 
\cite{Mor56,Kub57,Mor58,Yam67,Kaw73,Zub74,Kub78}, 
so it may be meaningful to mention the consistency of 
the non-equilibrium canonical distribution (\ref{GenerShearCanon2}) 
%discussed in the previous subsection \ref{CanonDistrShear} 
with the linear response formula for viscosity.
% in this subsection.    

   Using the notation (\ref{ShearAvera}) and 
the quantity $P_{xy}(\bfGamma)$ defined by Eq. (\ref{StresTenso0}),  
the viscosity $\eta$ is introduced as  

\begin{eqnarray}
   \eta \equiv - \lim_{\gamma\rightarrow 0} 
   \frac{\left\langle 
   P_{xy}(\bfGamma) \right\rangle_{\infty}}{\gamma}
\label{visco}\end{eqnarray}

\noindent Using the distribution $\tilde{f}(\bfGamma,t)$ 
given by Eq. (\ref{GenerShearCanon2}),  
the viscosity $\eta$ is represented as 

\begin{eqnarray}
   \eta = \beta \mathcal{V} 
   \int_{t_{0}}^{\infty} dt \;\; 
   \left\langle \tilde{P}_{xy}(\bfGamma,t) 
   P_{xy}(\bfGamma) 
   \right\rangle^{(eq)}
\label{GreenKuboVisco}\end{eqnarray}

\noindent where we introduced the notation 
$\langle X(\bfGamma) \rangle^{(eq)}$ as the 
equilibrium average of $X(\bfGamma)$ for any 
function $X(\bfGamma)$, 
namely $\langle X(\bfGamma) \rangle^{(eq)} \equiv 
\Xi^{(eq)-1} \int d\bfGamma X(\bfGamma) 
\exp\{-\beta H^{(ine)}(\bfGamma)\}$
with the equilibrium partition function 
$\Xi^{(eq)}$. 
   The derivation of Eq. (\ref{GreenKuboVisco}) 
is given in Appendix \ref{GreKubVisco}.   
   (In the same appendix \ref{GreKubVisco} we 
also discuss two kinds of nonlinear 
response formulas for 
$\left\langle P_{xy}(\bfGamma) \right\rangle_{\infty}$ 
with respect to the shear rate $\gamma$, 
one of which can be regarded as a natural generalization of 
Eq. (\ref{GreenKuboVisco}).)
   Here, it is important to note that 
   
\begin{eqnarray}
   \overline{P_{xy}(\bfGamma)} = 0, 
%   - \lim_{\gamma\rightarrow 0}
%   \frac{\overline{P_{xy}(\bfGamma)}}{\gamma} = 0,  
\label{zerothvisco}\end{eqnarray}
   
\noindent at any shear rate $\gamma$, 
as also shown in Appendix  \ref{GreKubVisco}, 
so that we obtain the equation 
$- \lim_{\gamma\rightarrow 0}
\overline{P_{xy}(\bfGamma)}/\gamma = 0$  
meaning that the distribution 
$f(\bfGamma)$ does not include information 
about the viscosity.
   Equation (\ref{GreenKuboVisco}) is the well known 
linear response formula for viscosity \cite{Kub78}.

%   The distribution function $f(\bfGamma) does not  
% include the information about the work to sustain the shear flow 
% and heat caused by it, and the viscosity is the quantity 
% concerning them. 
%   This is the reason why we should generalize the 
% distribution function $f(\bfGamma) so that 
% we can calculate the viscosity and it include 
% information of a relaxation process. 

% V^2 \int <Pxy*Pxy> = \lim_{t->infty} <Q^2>/(2t)

   The factor $\exp\{-\beta\gamma \mathcal{V} \int_{t_{0}}^{t} ds 
\tilde{P}_{xy}(\bfGamma,-s+2t_{0}) \}$
in the non-equilibrium canonical distribution 
(\ref{GenerShearCanon2}) gives the difference between 
the two distributions $f(\bfGamma)$ and $\tilde{f}(\bfGamma,t)$, 
and plays an essential role in the 
derivation of the linear response formula (\ref{GreenKuboVisco}) 
for viscosity. 
   It may be emphasized that this kind of factor can be 
derived from a different approach using 
the Sllod equation \cite{Eva90a,Mor85}. 
   The Sllod equations expresses the dynamics of 
the velocity corresponding to 
$\mathbf{v}^{(mov)}$, and has been used 
in many numerical and analytical works 
on shear flow systems \cite{Eva90a,Sar98}. 
   In the canonical distribution approach using 
the Sllod equations, the time-evolution of 
a canonical distribution under Sllod dynamics 
is considered, and it leads to the distribution 
evolving a time-integral of the shear stress, 
like the distribution (\ref{GenerShearCanon2}). 
   In Appendix \ref{SllodCanonical} we discuss briefly 
the relation between the 
Sllod dynamics approach and the Hamiltonian dynamics 
approach used here.  
   These two approaches give the same formula 
(\ref{GreenKuboVisco}) for the viscosity. 
   A difference between this approach and the 
approach discussed in this paper is that 
the Sllod dynamics approach is based on 
distributions of the type (\ref{CanonLocalEquib}), 
so that it does not take into account of 
the inertial force. 
   This make discussions of thermodynamic relations  
(for example, the first law of thermodynamics) 
rather more complicated than the approach used in  
this paper. 
   It may also be noted that the Sllod  
dynamics is different from the 
dynamics for $\mathbf{v}^{(mov)}$ 
from the Hamiltonian $H^{(mov)}(\bfGamma)$ 
in the moving frame $\mathcal{F}^{(mov)}$, and 
in the Sllod dynamics approach the distribution 
corresponding to $f(\bfGamma)$ is 
just an initial test distribution 
and cannot be interpreted as a steady 
distribution in the moving frame $\mathcal{F}^{(mov)}$ 
like the Hamiltonian dynamics approach 
discussed in this paper.

\subsection{Work Needed to Sustain Shear Flows and the House-Keeping Heat}
\label{ShearWork}

   Now we discuss further the information involved in the distribution 
$\tilde{f}(\bfGamma,t)$, which the canonical distribution 
$f(\bfGamma)$ does not have. 
   It is the information about the work required to sustain the 
steady shear flow. 

   First, the power $\langle\dot{W}\rangle_{t}$ 
to sustain the shear flow at time $t$ is estimated by 

\begin{eqnarray} 
   \langle\dot{W}\rangle_{t} 
   &\equiv& 
   \frac{d\; \langle 
   H^{(mov)} \rangle_{t}}{d t} \nonumber \\
   &=& 
   \frac{d\; \overline{\tilde{H}^{(mov)}(\bfGamma,t)}}{d t} 
   \nonumber \\
   &=& -\gamma \mathcal{V} \left\langle P_{xy}
   \right\rangle_{t} 
\label{PowerSustain}\end{eqnarray}

\noindent 
where $\tilde{H}^{(mov)}(\bfGamma,t)$ is defined 
as $\tilde{H}^{(mov)}(\bfGamma,t) 
\equiv \exp\{i\hat{L}^{(ine)}(t-t_{0})\} 
H^{(mov)}(\bfGamma)$. 
% and roughly speaking it 
%means the internal energy at time $t$ by looking 
%from the inertial frame. 
   Here, in order to derive Eq. (\ref{PowerSustain}) 
we used Eqs. 
%(\ref{ShearHamilMov}), (\ref{HelfaTimDer}), 
$i\hat{L}^{(ine)} H^{(ine)}(\bfGamma) = 0$ and 
$\partial \tilde{H}^{(mov)}(\bfGamma,t) / \partial t 
= \exp\{i\hat{L}^{(ine)}(t-t_{0})\} 
\mathcal{V} P_{xy}(\bfGamma)$.

%addition
\vspace{0.2cm}
%---------------------------------------------------------------------
\begin{figure}[!htb]
\vspfigA
\includegraphics[width=0.45\textwidth]{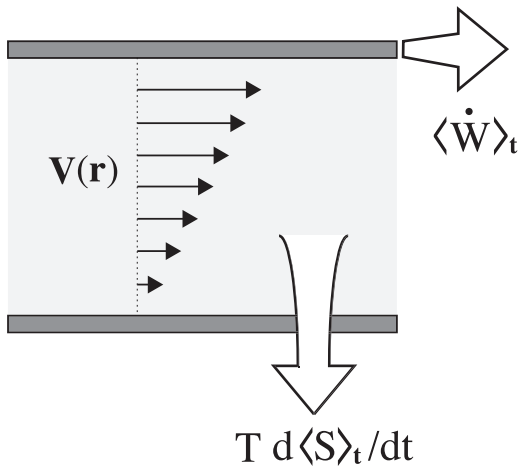}
\vspace{-0.5cm}
\caption{
      Schematic illustration of the power $\langle\dot{W}\rangle_{t}$ 
   required to sustain the shear flow and the house-keeping heat 
   $T [d\langle S \rangle_{t}/dt]$. 
      In this illustration the power $\langle\dot{W}\rangle_{t}$ 
   is represented as the power to move the upper boundary of 
   the shearing system.  
      This power supplies the energy to sustain the shear flow state, 
   which is eliminated from the system as the house-keeping heat. 
   }
\vspfigB
\label{fig0WorkHeat}\end{figure}  
%-------------------------------------------------------------------- 

   In steady flow systems, the energy added 
to the system by the work needed to sustain the flow 
must be eliminated from the system as heat  \cite{Eva83}. 
   This type of heat is called the "house-keeping heat", 
and its special role has been emphasized 
in the construction of a non-equilibrium steady state 
thermodynamics \cite{Ono98,Hat01}. 
%addition
   Fig. \ref{fig0WorkHeat} is a schematic illustration of 
the power $\langle\dot{W}\rangle_{t}$ required to sustain 
the shear flow and the house-keeping heat.
   Now we estimate this house-keeping heat from 
the non-equilibrium canonical distribution approach. 
   For this purpose we introduce the non-equilibrium entropy 
$\langle S \rangle_{t}$ as the ensemble average of $S(\bfGamma)$ by 
the distribution $\tilde{f}(\bfGamma,t)$.  
%\begin{eqnarray}
%   \tilde{S}(t) 
%   \equiv \langle S(\bfGamma) \rangle_{t}
%\label{GenEntro}\end{eqnarray}
%\noindent 
   Note that $S(\bfGamma)$ is defined by Eq. (\ref{ObserEntro}), 
and was already used as the observable corresponding 
to the entropy in Eq. (\ref{RotEntropy3}). 
   A similar kind of entropy to $\langle S \rangle_{t}$
%Eq. (\ref{GenEntro}) 
was used in Refs. \cite{Mor56,Mor58} for a 
different form of the distribution $f(\bfGamma)$. 
   Using Eq. (\ref{GenerShearCanon2}) 
%and (\ref{GenEntro}) 
%the Hermiticity of the Liouville operator $\hat{L}^{(ine)}$ 
the entropy $\langle S \rangle_{t}$ 
is represented as    

\begin{eqnarray}
   \langle S \rangle_{t} &=&
   - \int d\bfGamma f(\bfGamma) 
   \ln \left\{\tilde{f}(\bfGamma,-t+2t_{0}) \right\} 
   \nonumber \\
   &=& \overline{S} + \beta
      \gamma \mathcal{V} 
      \int_{t_{0}}^{-t+2t_{0}} ds 
      \overline{\tilde{P}_{xy}(\bfGamma,-s+2t_{0})}
   \nonumber \\
   &=& \overline{S} - \beta
      \gamma \mathcal{V} 
      \int_{t_{0}}^{t} ds 
	  \left\langle P_{xy} 
   \right\rangle_{s} 
%      \overline{\tilde{P}_{xy}(\bfGamma,s)}
\label{GenEntro2}\end{eqnarray}

\noindent Therefore the house-keeping heat 
$T [d\langle S \rangle_{t}/dt]$ at time $t$ is given by 

\begin{eqnarray}
   T\frac{d\langle S \rangle_{t}}{dt} &=& - 
      \gamma \mathcal{V} 
	  \left\langle P_{xy} 
   \right\rangle_{t} 
%      \overline{\tilde{P}_{xy}(\bfGamma,t)} 
   \label{HouseKeepHeat1} \\
   &=& \langle\dot{W}\rangle_{t} .
\label{HouseKeepHeat2}\end{eqnarray}

\noindent This is the balance equation to show that 
the power needed to sustain the shear flow must be equal 
to the house-keeping heat. 
%meaning that the energy given in the system as the 
%work to sustain the shear flow is eliminated from the system 
%as the house-keeping heat. 
   Using Eqs. (\ref{visco}), 
(\ref{HouseKeepHeat1}) and (\ref{HouseKeepHeat2}),  
the house-keeping heat and the ensemble averaged power 
$\langle\dot{W}\rangle_{t}$ to sustain the shear flow is connected 
to the viscosity $\eta$ as $T [d\langle S \rangle_{t}/dt] 
= \langle\dot{W}\rangle_{t} = \mathcal{V} \eta \gamma^{2} 
+ \mathcal{O}(\gamma^{3})$. 

   The entropy $\langle S \rangle_{t}$ 
%defined by Eq. (\ref{GenEntro}) 
satisfies the inequality 

\begin{eqnarray}
   \langle S \rangle_{t} \geq \langle S \rangle_{t_{0}} = \overline{S}
\label{SecondLaw}\end{eqnarray}

\noindent at any time $t \; (>t_{0})$. 
   The detail of the derivation of the inequality 
(\ref{SecondLaw}) is given in Appendix \ref{AppenSecondLaw}. 
   Noting that $\langle S \rangle_{t} - \langle S \rangle_{t_{0}} 
= \int_{t_{0}}^{t} dt [d \langle S \rangle_{t} /dt]$ and assuming  
that $d \langle S \rangle_{t} / dt$ is time-independent in 
a steady state, we obtain 

\begin{eqnarray}
   \frac{d\langle S \rangle_{t}}{dt} \geq 0
\label{SecondLaw2}\end{eqnarray}

\noindent This is the expression of the second law of thermodynamics 
in the non-equilibrium canonical distribution approach.
   The inequality (\ref{SecondLaw2}) means simply that 
the shear flow system produces a positive house-keeping 
heat constantly in time.  
   In this sense, the total entropy production 
$\langle S \rangle_{t} - \langle S \rangle_{t_{0}}$ diverges 
as time $t$ go to infinity, 
because the total amount of heat produced by the steady visco-elastic 
shear flow in the infinite time interval is infinite \cite{noteIV1}. 
%, as emphasized in Ref. \cite{Hat01}. 
   In other words, the system is kept as a non-equilibrium steady 
state by discharging an amount of entropy constantly, which is 
transferred from the external work. 
   Therefore the inequality (\ref{SecondLaw2}) must be distinguished 
from another type of second law of thermodynamics meaning 
that an entropy increases in time and approaches to a stable value 
in a relaxation process. 
   This type of the second law of thermodynamics, or 
the thermodynamical stability condition, will be discussed 
in Sec. \ref{StabiShear}. 
   By combining the inequality (\ref{SecondLaw2}) and $T>0$ 
with Eqs. (\ref{HouseKeepHeat1}) and (\ref{HouseKeepHeat2}) 
we have 

\begin{eqnarray}
   \frac{\langle\dot{W}\rangle_{\infty}}{\mathcal{V}}
   = - \gamma \left\langle P_{xy}(\bfGamma) 
   \right\rangle_{\infty} \geq 0.
\label{positiveheat}\end{eqnarray}

\noindent Namely the averaged power 
$\langle\dot{W}\rangle_{\infty}$ 
needed to sustain the shear flow must be positive (or zero). 
   This is one of the results,  
which can be checked by numerical simulation, 
as will actually shown in Sec. \ref{HouseKeepingHeatNumeric}. 
   It may be noted that the inequality (\ref{positiveheat}) 
implies the non-negativity of the viscosity $\eta$ as a special case.

%%%%%%%%%%%%%%%%%%%%%%%%%%%%%%%%%%%%%%%%%%%%%%%%%%%%%%%%%%%%%%%%%%%%%%)

\section{Thermodynamics for Shear Flow Systems}
\label{ShearFlowsTherm}

   As discussed in Sec. \ref{ShearWork}, the 
factor $\exp\{\gamma \mathcal{V} 
\int_{t_{0}}^{t} ds \tilde{P}_{xy}(\bfGamma,-s+2t_{0})
\}$, which gives the difference between the distribution 
$\tilde{f}(\bfGamma,t)$ and the canonical distribution 
$f(\bfGamma)$,  
%and plays an essential role to calculate the 
%transport coefficient like the viscosity, 
includes the effect of the work needed to 
sustain the shear flow, or the house-keeping heat. 
   On the other hand, the expression for 
the first law of thermodynamics proposed by Evans and Hanley 
does not include this effect 
(Otherwise it must include time-dependent terms for 
 the sustaining work and the house-keeping heat.). 
   Moreover, Ref. \cite{Hat01} emphasized that we must subtract 
the contribution of the house-keeping heat from the entropy 
in order to obtain an expression for 
the first law of thermodynamics for steady states. 
   Equation (\ref{GenEntro2}) implies that the 
entropy minus the contribution from the house-keeping heat 
is given by $\langle S \rangle_{t} - 
\int_{t_{0}}^{t} ds [d\langle S \rangle_{s}/ds]
=\langle S \rangle_{t}+ \beta\gamma \mathcal{V} 
\int_{t_{0}}^{t} ds \langle P_{xy}(\bfGamma) 
\rangle_{s}
%\overline{\tilde{P}_{xy}(\bfGamma,s)} 
= \overline{S}$, which is the entropy 
defined through the canonical distribution $f(\bfGamma)$, 
not through the distribution $\tilde{f}(\bfGamma,t)$. 
   For these reasons, 
(although it may not be impossible to construct 
a non-equilibrium steady state thermodynamics explicitly 
including the effect of the house-keeping heat) 
in this section we construct 
a shear flow thermodynamics based on the 
canonical distribution  $f(\bfGamma)$ 
excluding the effect of the house-keeping heat, and show that it  
is consistent with the Evans-Hanley thermodynamics.

%---------------------------------------------------------------------
\subsection{First Law of Thermodynamics for Shear Flows}
\label{FirstLawShear}

   As the first thermodynamic property of the shear flow system, 
we consider the first law of thermodynamics.  
   Its derivation is quite similar to that for the 
rotating system discussed in Sec. \ref{ThermoRotat}, 
so it is given rather briefly. 

   In the shear flow system, the entropy $\overline{S}$ is given by 
Eq. (\ref{RotEntropy2}) or 

\begin{eqnarray}
   \overline{S} 
%      &=& \ln\Xi + \beta 
%      \overline{H^{(mov)}} 
%      \label{ShearEntro1} \\
%      &=& 
	  =\ln\Xi + \beta \left[ 
      \overline{H^{(ine)}} - \gamma \overline{Q} \right] 
\label{ShearEntro2}\end{eqnarray}

\noindent using the canonical distribution 
(\ref{ShearCanon}). 
%This entropy $\overline{S}$ does not include the term caused by 
%the house-keeping heat. 
   As in the rotating system, 
the free energy $F^{(mov)}$ in the moving frame 
$\mathcal{F}^{(mov)}$ is defined by Eq. ({\ref{RotFreEneMov0}), 
and is connected to the partition function $\Xi$ by 
Eq. ({\ref{RotFreEneMov1}). 
   (Here, it may  be noted that the free energy $F^{(mov)}$ 
can also be expressed as $F^{(mov)} 
= \langle H^{(mov)}(\bfGamma) \rangle_{t} - T \langle S \rangle_{t}$ 
using the averaged energy 
$\langle H^{(mov)}(\bfGamma) \rangle_{t}$ 
and entropy $\langle S \rangle_{t}$ related 
to the distribution $\tilde{f}(\bfGamma,t)$ which 
includes information about the house-keeping heat.)
    Using this free energy $F^{(mov)}$ we obtain Eqs. 
 (\ref{RotDerFreEneMovTem}) and 

\begin{eqnarray}
   \frac{\partial \left[\beta F^{(mov)} \right]}{\partial \gamma} 
   = - \frac{\partial \ln\Xi}{\partial \gamma} 
   = - \beta\overline{Q}
\label{ShearDerFreEneMovAng}\end{eqnarray}

\noindent which are equivalent to  
$d [\beta F^{(mov)} ] = \overline{H^{(mov)}} d\beta 
- \beta \overline{Q} d\gamma $. 
   Therefore we obtain 

\begin{eqnarray}
   d F^{(mov)} 
   = - \overline{S} dT 
   - \overline{Q} d\gamma ,
\label{SearFirstLaw1}\end{eqnarray}
   
\noindent noting Eq. ({\ref{RotFreEneMov0}).
   Using the relations 
$F^{(ine)} = F^{(mov)} - \gamma \overline{Q}$, 
$\overline{H^{(mov)}} = F^{(mov)} + T \overline{S}$ 
and $\overline{H^{(ine)}} = F^{(ine)} + T \overline{S}$ 
we also obtain 
   
\begin{eqnarray}   
   d F^{(ine)} = - \overline{S} dT 
+ \gamma d\overline{Q}  ,
\label{SearFirstLaw2}\end{eqnarray}
\begin{eqnarray}
   d \overline{H^{(mov)}}
   = T d\overline{S} 
   - \overline{Q} d\gamma ,
\label{SearFirstLaw3}\end{eqnarray}
\begin{eqnarray}
   d \overline{H^{(ine)}}
   = T d\overline{S} 
     +\gamma d\overline{Q} .
\label{SearFirstLaw4}\end{eqnarray}
   
\noindent The two energies 
$\overline{H^{(ine)}}$ and $\overline{H^{(mov)}}$
in the different frames 
are connected by a Legendre transformation, namely  
$\overline{H^{(ine)}}=\overline{H^{(mov)}} 
- \gamma \;
\partial \overline{H^{(mov)}}/ \partial \gamma|_{\overline{S}}$, 
as well as the two free energies in the different frames.
   It is clear that the Evans-Hanley expression 
(\ref{EvansHanley}) 
for the first law of shear flow thermodynamics is 
the relation (\ref{SearFirstLaw3}) where 
$\mathcal{E}=\overline{H^{(mov)}}$
and $\xi = - \overline{Q}$. 
%   It must be emphasized that for example, 
%the work $-\overline{Q}d\gamma$ by a change of the shear rate 
%$\gamma$ in the relation (\ref{SearFirstLaw3}) 
%does not include the work to sustain the shear flow steadily, 
%because the distribution $f(\bfGamma)$ does not include 
%such an effect, 
%as already discussed in the preceding section. 

%---------------------------------------------------------------------
\subsection{Thermodynamic Stability Conditions for Shear Flows} 
\label{StabiShear}

   As the second thermodynamical property of the shear flow 
system, although we omitted to discuss it in Sec. \ref{ThermoRotat} 
for the rotating system, we consider a stability condition 
for shear flow \cite{note3}.  
%based on the fact that the entropy $\overline{S}$ 
%should take a local minimum value in a steady state. 

   We consider a small part $\mathcal{A}$ of the 
macroscopic shear flow system, in which averages of energy, 
entropy and Helfand's moment of viscosity  
in the inertial frame $\mathcal{F}^{(ine)}$ 
are given by $\overline{H^{(ine)}}$, 
$\overline{S}$ and $\overline{Q}$, respectively. 
   The other part $\mathcal{R}$ of the system, 
which is much bigger than the system $\mathcal{A}$ 
and is called 
the "environment" or "reservoir", has the thermodynamical 
values $T_{0}$, $\overline{S}_{0}$ and $\gamma_{0}$ of the 
temperature, the entropy and 
the shear rate, respectively.  
   Now, we consider moving an infinitesimal amount of 
energy as heat $-T_{0}d\overline{S}_{0}$ from the reservoir 
$\mathcal{R}$ into the small system $\mathcal{A}$. 
   In this process the total entropy must increase: 
$d\overline{S}+d\overline{S}_{0}\geq 0$. 
   By combining this inequality with the first law of thermodynamics 
$d\overline{H^{(ine)}} 
= -T_{0}d\overline{S}_{0} + \gamma_{0}d\overline{Q}$ 
based on Eq. (\ref{SearFirstLaw4}), 
we have 
%
%\begin{eqnarray}
$       
   d \overline{H^{(ine)}} - T_{0}d\overline{S} 
   -\gamma_{0} d\overline{Q} 
   = 
   d (\overline{H^{(ine)}} - T_{0} \overline{S} 
   -\gamma_{0} \overline{Q})   \leq 0
$,
%\label{ShearStabi1}\end{eqnarray} 
%
%\noindent 
using the fact that the reservoir $\mathcal{R}$ is so big 
that $T_{0}$ and $\gamma_{0}$ do not change in 
this process. 
   This inequality means that the quantity 
$\overline{H^{(ine)}} - T_{0} \overline{S} -\gamma_{0} \overline{Q}$ 
always decreases and reaches a minimum at a stable state. 
   In other words, if we force a change to the values of 
$\overline{H^{(ine)}}$, $\overline{S}$ and $\overline{Q}$ at 
the stable point by $\delta\overline{H^{(ine)}}$, 
$\delta\overline{S}$ and $\delta\overline{Q}$, respectively,   
then the inequality 
$       
   \delta \overline{H^{(ine)}} - T_{0}\delta\overline{S} 
   -\gamma_{0} \delta\overline{Q} \geq 0
$
must be satisfied as the stability condition for the shear 
flow system. 
   This simply leads to 

\begin{eqnarray}
    \delta^{2} \overline{H^{(ine)}}
   %(\overline{S},\overline{Q}) 
   \geq 0 
\label{ShearStabi2}\end{eqnarray} 

\noindent for any infinitesimal 
deviations $\delta \overline{S}$ and $\delta \overline{Q}$. 
   By a well known technique used in 
thermodynamics (See, for example,  Ref. \cite{Lan59}
or Appendix \ref{ShearStabiCondi}.), 
the condition (\ref{ShearStabi2}) is equivalent to  

\begin{eqnarray}
   \left.\frac{\partial \overline{S}}
   {\partial T}\right|_{\overline{Q}} > 0, 
\label{ShearStabi3a}\end{eqnarray} 
\begin{eqnarray}
   \left.\frac{\partial\overline{Q}}{\partial \gamma}\right|_{T} 
   > 0 .
\label{ShearStabi3b}\end{eqnarray} 

\noindent The condition (\ref{ShearStabi3a}) simply means that 
the specific heat at constant $\overline{Q}$ is always positive 
at a positive temperature $T$. 
   To understand the condition (\ref{ShearStabi3b}) 
%\cite{noteSta}, 
we note 
 
\begin{eqnarray}
   \left.\frac{\partial\overline{Q}}{\partial \gamma}\right|_{T} 
   = \beta \left(\overline{Q^{2}}- \overline{Q}^{2}\right)
\label{ShearStabi3c}\end{eqnarray} 

\noindent as shown in Appendix \ref{ShearStabiCondi}. 
		   Therefore, combining Eq. (\ref{ShearStabi3c}) 
with the inequality (\ref{ShearStabi3b}) 
we obtain 

\begin{eqnarray}
   \overline{Q^{2}}- \overline{Q}^{2} > 0 .
\label{ShearStabi3d}\end{eqnarray}

\noindent Namely, the stability condition (\ref{ShearStabi3c}) 
means the positivity of the correlation function for 
Helfand's moment $Q(\bfGamma)$ of viscosity.

%
%\bibitem{noteSta} 
Based on Eq. (\ref{EvansHanley}), 
Evans and Hanley claimed the inequality 
$\partial \xi/\partial \gamma|_{T} > 0$ as a stability condition 
for shear flows \cite{Han82,Eva80b}. 
   This inequality is incompatible with the inequality 
(\ref{ShearStabi3b}) in the case of $\xi=-\overline{Q}$. 
   This difference comes basically from the fact that 
they discussed a thermodynamic stability 
condition using the energy $\overline{H^{(mov)}}$, 
whereas we discussed it using the energy $\overline{H^{(ine)}}$. 
   Obviously the correlation function of $Q$ cannot be negative 
because of $\overline{Q^{2}}- \overline{Q}^{2} = 
\overline{(Q - \overline{Q})^{2}} \geq 0$, 
so noting Eq. (\ref{ShearStabi3c}) we cannot justify 
the stability condition  
claimed by Evans and Hanley
in the canonical distribution approach.

%---------------------------------------------------------------------
\subsection{Relations Between Canonical Averages} 
%\subsection{Relation Between Time and Canonical Averages} 
\label{twoaverage}

   So far we have introduced two types of canonical 
average  $\overline{X}$ and $\langle X \rangle_{t}$,
and in Sec. \ref{numeritest} we introduce the usual
time average.
   It is very important to distinguish between these
averages.
   The thermodynamic relations discussed 
in Secs. \ref{FirstLawShear} and \ref{StabiShear}
are the relations for the ensemble average $\overline{X}$ 
of observable $X(\bfGamma)$ using the canonical distribution 
$f(\bfGamma)$. 
   On the other hand in numerical simulations 
using the Sllod equations with an isokinetic thermostat 
(as in Sec. \ref{numeritest}), the 
values obtained are the mixed ensemble-time average 
$\langle X \rangle_{t}$ for the distribution 
$\tilde{f}(\bfGamma,t)$ in the limit $t\rightarrow \infty$. 
   Therefore it is important to obtain an explicit 
relation between these two different ensemble averages. 

   For any function $X(\bfGamma)$ 
the relation between 
the two ensemble averages $\overline{X}$ and 
$\langle X \rangle_{\infty}$ is 

\begin{eqnarray}
   && \langle X \rangle_{\infty} \nonumber \\
   &&= \overline{X} 
      - \beta \gamma \mathcal{V} \int_{t_{0}}^{\infty} dt \; 
      \overline{
      \left[\tilde{X}(\bfGamma,t) - \overline{X}\right]
      \Bigl[P_{xy}(\bfGamma) -  \overline{P_{xy}} \; \Bigr]
      } 
   \nonumber \\
\label{TwoAvera}\end{eqnarray} 

\noindent where $\tilde{X}(\bfGamma,t)\equiv 
\exp\{i\hat{L}^{(ine)}(t-t_{0})\}X(\bfGamma)$. 
   The derivation of Eq. (\ref{TwoAvera}) is given in 
Appendix \ref{TwoAveRel}. 
   A similar equation for 
the canonical distribution approach using the Sllod 
equations is shown in Ref. \cite{Eva87a}. 

   From relation (\ref{TwoAvera}), 
if the fluctuation $\tilde{X}(\bfGamma,t) - \overline{X}$ 
of $X$ is weakly correlated to the shear stress 
$P_{xy}(\bfGamma)$, then the ensemble average $\overline{X}$ 
can be nicely approximated by the average $\langle X \rangle_{\infty}$. 
   However one must notice that the justification for 
such an approximation strongly depends on the quantity $X$ we consider.
   A typical example is the case of $X=P_{xy}(\bfGamma)$, 
in which we must not neglect the second term on the right-hand side 
of Eq. (\ref{TwoAvera}), 
because in this case the first term on the right-hand side 
of Eq. (\ref{TwoAvera}) is zero,   
namely $\overline{P_{xy}}=0$, as shown in Appendix \ref{GreKubVisco}. 
   One should also notice that the second term 
in Eq. (\ref{TwoAvera}) is small near equilibrium, 
because it includes the 
non-equilibrium parameter $\gamma$ as a factor.

%---------------------------------------------------------------------
%\subsection{Work in a Thermodynamic Cycle Engine}

%%%%%%%%%%%%%%%%%%%%%%%%%%%%%%%%%%%%%%%%%%%%%%%%%%%%%%%%%%%%%%%%%%%%%%

\section{Numerical Simulations of Shear Flow}
\label{numeritest} 

   In this section we show numerical results for some 
quantities which have appeared 
%in discussion of the shear flow thermodynamics 
in the preceding sections 
\ref{ShearFlowCano} and \ref{ShearFlowsTherm}, 
and check the results obtained there. 

   For this numerical calculation we use a two-dimensional 
square system consisting of $N$ particles with 
a square shape and side length $L(=\sqrt{\mathcal{V}})$. 
%$0 \leq q_{jx}\leq L$ and $0 \leq q_{jy}\leq L$. 
   The particle-particle interaction is given by the 
isotropic soft-core pair-potential 

\begin{eqnarray}
   \phi(r) = \left\{
   \begin{array}{ll}
      \kappa\left(\frac{1}{r^{12}} 
         - \frac{1}{r_{0}^{12}}\right) 
               & \mbox{in} \;\; r <    r_{0}        \\
      0        & \mbox{in} \;\; r \geq r_{0} 
   \end{array}
   \right.
\label{simulpoten}\end{eqnarray} 

\noindent with a positive constants $r_{0}=1.5$ and 
$\kappa=1$. 
%as a function of the distance $r$ 
%between two particles. 
%and it is truncated at $r=\beta$. 
   The particle number density 
$\rho\equiv N/\mathcal{V}$ is $0.8$. 
   The mass $m$ of the particle and the kinetic temperature $T$ 
are both chosen as $1$. 
   The number of particles is $N=450$, 
except in Sec. \ref{remarksimul} in which the $N$-dependence 
of a quantity will be discussed. 
   We use the Sllod equations with Lees-Edwards boundary 
conditions and the isokinetic thermostat so that 
the kinetic temperature (given by Eq. 
(\ref{KinTempe})) is kept constant \cite{Eva90a}. 
   (This dynamics is explained in Appendix 
\ref{SllodCanonical} more explicitly.)
   A predictor-corrector method \cite{Gea71} 
of 4-th order 
is used to carry out these numerical simulations 
with time step of $\Delta t = 0.001$. 
   In this algorithm the sum of the "thermal momentum" 
$\tilde{\mathbf{p}}_{j}\equiv \mathbf{p}_{j} 
- m \mathbf{V}(\mathbf{q}_{j})$
is zero in the both coordinate directions.    

   We use the notation $\langle X \rangle$ for 
the time-averaged value of any quantity $X$ 
given by this numerical simulation. 
   To calculate this average we used data over more than 
$10^6$ time steps 
omitting the first $10^4$ time steps. 
   (We checked that $10^4$ time steps is much longer 
than the relaxation time of the time-correlation function 
for the thermal momentum.)
   This should correspond to the ensemble average 
$\langle X \rangle_{\infty}$ used so far. 
   This is supposed by the fact that 
we can calculate the viscosity 
from the time-average 
$\langle P_{xy} \rangle$ in this simulation, 
based on Eq. (\ref{visco}) assuming 
$\langle P_{xy} \rangle_{\infty} = \langle P_{xy} \rangle$. 
   We calculate $\gamma$-dependences 
of three quantities: $\langle P_{xy} \rangle$, 
$\langle Q \rangle$ and $\langle Q^{2} \rangle$. 
   We use $\langle P_{xy} \rangle$ to discuss 
the power to sustain the flow and 
the house-keeping heat given by Eq. (\ref{HouseKeepHeat1}). 
   The quantities $\langle Q \rangle$ and 
$\langle Q^{2} \rangle$ are used to discuss the behavior of 
Helfand's moment of viscosity and the thermodynamic 
stability condition (\ref{ShearStabi3b}). 
   Here, it is assumed that the quantities 
$Q$ and $P_{xy}$ are not strongly correlated 
with each other because of the relation (\ref{HelfaTimDer}), 
and in the case of $X=Q$ or $Q^{2}$ the second term 
of the right-hand side of Eq. (\ref{TwoAvera}) 
may be small compared to the first term in the 
small shear rate case. 
%   (Our numerical calculations support this assumption.) 
   This implies that the behavior of the 
time-averages of Helfand's moment of viscosity 
and its correlation function are not so different from 
the ones for $\overline{Q}$ 
and $\overline{Q^{2}}-\overline{Q}^{2}$, respectively, 
near equilibrium. 

%we use the same notation 
%$\overline{X}$ for the time-average of a quantity $X$ 
%in the case of $X=Q$ or $Q^{2}$ 
%as the ensemble average used in the preceding sections, 
%although it is not so obvious 
%in a non-equilibrium molecular simulation 
%as explained in the end of this section. 
%   Such an assumption is not so trivial 
%in non-equilibrium systems such as shear flows, 
%but has been used in many calculation of transport 
%coefficients in non-equilibrium molecular dynamics. 

%---------------------------------------------------------------------
\subsection{Work Needed to Sustain the Shear Flow}
\label{HouseKeepingHeatNumeric}

   The first numerical result is for 
the power $\dot{W}/\mathcal{V}$ per unit volume 
($\mathcal{V}= 562.499 \cdots$) for the work required to 
sustain the shear flow, 
or equivalently in a quantitative sense, the house-keeping heat 
per unit volume. 
%\mathcal{V}= 562.4999916 
   It is given as the time-average of 
$\dot{W}/\mathcal{V} \equiv -\gamma P_{xy}(\bfGamma)$, 
based on Eqs. (\ref{HouseKeepHeat1}) 
and (\ref{HouseKeepHeat2}). 
   Figure \ref{fig1heat} shows the shear rate $\gamma$ dependence 
of the time-averaged power $\langle \dot{W} \rangle /\mathcal{V}$ 
per volume.  
   The inset is the same graph except showing it 
in a wider shear rate region. 
   (Note that we used a linear-log scale in this inset, 
whereas we use a linear-linear scale for the main figure.)
   Following from the inequality (\ref{positiveheat}), 
the power to sustain the flow shown in Fig. \ref{fig1heat} 
is always positive (or zero).

%---------------------------------------------------------------------
\begin{figure}[!htb]
\vspfigA
\includegraphics[width=\widthfigA]{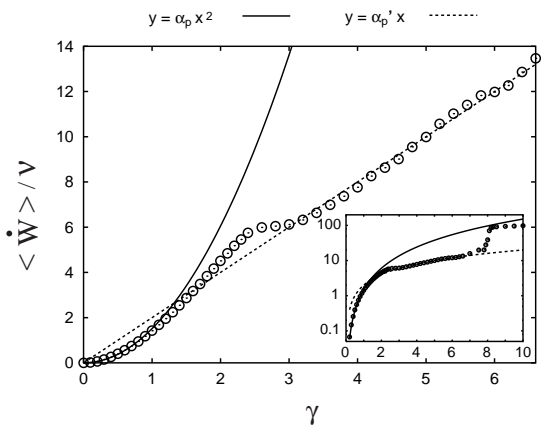}
\caption{
      The averaged power $\langle \dot{W} \rangle /\mathcal{V}$ 
   divided by volume $\mathcal{V}$ needed to sustain the shear flow,
   as a function of shear rate $\gamma$ as a linear-linear plot. 
      The solid line is the fit to a quadratic function 
   and the dashed line is the fit to a linear function (valid for 
   $\gamma > 3$). 
      Inset: the same graph as a linear-log plot including 
   a wider range of $\gamma$. 
   }
\vspfigB
\label{fig1heat}\end{figure}  
%-------------------------------------------------------------------- 

   It may be noted that the averaged power 
$\langle \dot{W} \rangle$ needed to sustain the shear flow
should be an even function of $\gamma$, because 
it should be invariant under 
a change of sign of the shear rate $\gamma$. 
   In Fig. \ref{fig1heat} 
we fitted the numerical data to a quadratic function 
$y=\alpha_{P} x^{2}$ with the fitting parameter 
$\alpha_{P}= 1.51309$. 
%C1              = 1.51309          +/- 0.01421      (0.9392%)
   Near equilibrium $\gamma < 0.5$, 
the graph is nicely fitted by this quadratic function. 
%the coefficient $-\alpha_{P}$ gives the viscosity at equilibrium. 

   As the shear rate increases, a region in which 
the value of $\langle P_{xy}(\bfGamma)\rangle$ is almost 
independent of $\gamma$, (namely the region fitted by 
a linear function $y=\alpha_{P}' x$ with the fitting parameter 
$\alpha_{P}'= 2.00005$), appears \cite{Woo84}, 
%B1              = 2.00005          +/- 0.008683     (0.4341%)
and after that the region the string phase appears \cite{Erp84,Woo84,Hey85} 
(the region $\gamma > 8$ approximately in the inset 
to Fig. \ref{fig1heat}). 
   The string phase can be checked not only by the string-type 
arrangement of particle positions but also by strong 
time-oscillating behavior of the time-correlation functions 
for quantities such as the potential energy, 
the shear stress and so on \cite{Hey85}. 
   
   For isokinetic thermostat, the house-keeping heat 
is also given as the time-average of the 
thermostat term (the term $- \alpha(t) \mathbf{p}'_{j}(t)$ 
explained in Appendix \ref{SllodCanonical}). 
   We checked numerically that this quantity is 
equal to the time-average of $-\gamma P_{xy}(\bfGamma)$.

%---------------------------------------------------------------------
\subsection{Helfand's Moment of Viscosity}

   As a second numerical result, 
Fig. \ref{fig2Q} shows the graph of the time-average 
$\langle Q\rangle/\mathcal{V}$ of the  Helfand's moment  
of viscosity per unit volume 
$\mathcal{V}$ as a function 
of shear rate $\gamma$. 
%\mathcal{V}= 562.4999916 
   It (almost) takes the value $0$ 
at the equilibrium $\gamma=0$, 
and increases linearly as a function of $\gamma$. 
   In this figure we also give a fit to a 
linear function $y=\alpha_{Q} \gamma$, with the 
parameter value $\alpha_{Q} = 150.043$. 
%A               = 150.043          +/- 0.3999       (0.2665%)

%---------------------------------------------------------------------
\begin{figure}[!htb]
\vspfigA
\includegraphics[width=\widthfigA]{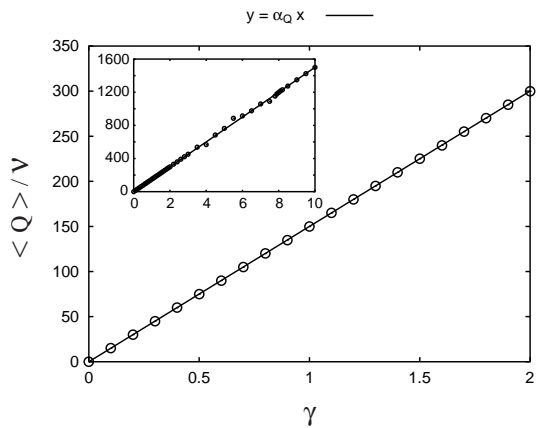}
\caption{
      Time-average $\langle Q\rangle/\mathcal{V}$ 
   of Helfand's moment 
   of viscosity divided by volume $\mathcal{V}$ 
   as a function of shear rate $\gamma$. 
      The solid line is the fit to a linear function. 
      Inset: the same graph including a wider range of $\gamma$. 
   }
\vspfigB
\label{fig2Q}\end{figure}  
%-------------------------------------------------------------------- 

   To explain this linear behavior for Helfand's moment 
of viscosity as a function of shear rate, we simply note that 

\begin{eqnarray}
   \left\langle Q \right\rangle 
   = \sum_{j=1}^{N}  \left\langle q_{jy} \tilde{p}_{jx} \right\rangle
   + \gamma \sum_{j=1}^{N}  \left\langle q_{jy}^{2} \right\rangle
\label{Helfa2}\end{eqnarray}

\noindent with the $x$-component 
$\tilde{p}_{jx}\equiv p_{jx} - \gamma q_{jy}$ 
of the thermal momentum of the $j$-th particle. 
   Our numerical calculations show that the value of 
the first term of the 
right-hand side of Eq. (\ref{Helfa2}) is extremely 
small (or zero) compared to the value of its second term,  
namely $\sum_{j=1}^{N}  \langle q_{jy} \tilde{p}_{jx} 
\rangle \approx 0$. 
   Moreover, using a homogeneous continuum assumption for the fluid, 
the value of the quantity 
$\sum_{j=1}^{N}  \langle q_{jy}^{2}\rangle$ appearing 
in the second term of the right-hand side of Eq. (\ref{Helfa2}) 
can be estimated as 
$\sum_{j=1}^{N}  \langle q_{jy}^{2}\rangle \approx 
NL^{-1}\int_{0}^{L}dy y^{2} = N L^{2}/3$. 
   These estimations lead to 
$\langle Q\rangle /\mathcal{V}\approx(N/3)\gamma = 150\gamma$, 
which explains the value of the fitting parameter $\alpha_{Q}$. 

   The time-averaged Helfand's moment 
$\langle Q\rangle$ of viscosity should be at least an odd function 
of shear rate $\gamma$, because the infinitesimal 
deviation $\gamma d \overline{Q}$ giving the energy 
change $d \overline{H^{(ine)}}$
in the inertial frame by Eq. (\ref{SearFirstLaw4}) 
must be invariant under the change of sign 
of the shear rate. 
   It may be noted that this linear dependence for the 
time-average of Helfand's moment $Q$ 
of viscosity with respect to shear rate is satisfied 
not only in the near equilibrium region but also even in the 
string phase region, shown in the inset to Fig. \ref{fig2Q}, 
possibly because the Sllod equations are a homogenous shear algorithm. 

   It may be noted that in our simulations Helfand's moment 
of viscosity can be generally changed discontinuously in time, 
when a particle steps over a boundary in the direction orthogonal to  
the global shear flow. 
   However it should be a small boundary effect which can be 
neglected in the thermodynamic limit $N\rightarrow\infty$ 
and $\rho=const$, 
and our numerical calculations gave a good convergence for the 
long time-average of Helfand's moment of viscosity.

%---------------------------------------------------------------------
\subsection{Correlation Function for Helfand's Moment of Viscosity}
\label{CorrelationFunctionHelfand}

   As the last example, Fig. \ref{fig3Qcor} 
shows the shear rate dependence of the correlation function 
$(\langle Q^{2}\rangle - \langle Q \rangle^{2})/\mathcal{V}$ 
of Helfand's moment of viscosity 
divided by the volume $\mathcal{V}$. 
   This figure shows that this correlation function 
is always positive 
at least for $\gamma<10$, following 
the thermodynamic stability condition (\ref{ShearStabi3b}).

%---------------------------------------------------------------------
\begin{figure}[!htb]
\vspfigA
\includegraphics[width=\widthfigA]{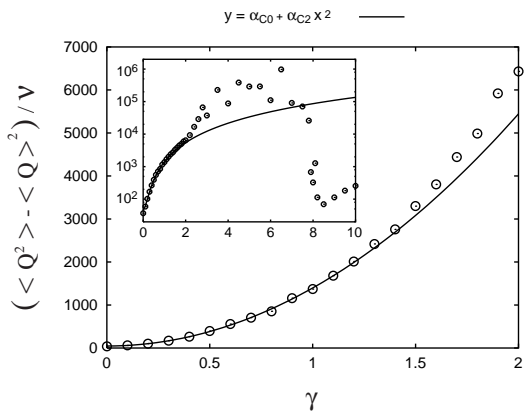}
\caption{
      Correlation function 
   $(\langle Q^{2}\rangle - \langle Q \rangle^{2})/\mathcal{V}$ of 
   Helfand's moment of viscosity divided by volume 
   $\mathcal{V}$
   as a function of shear rate $\gamma$ in a linear-linear plot. 
      The solid line is the fit to a quadratic function.
      The inset: the same graph except for that it includes 
   a wider region of  $\gamma$ and is a linear-log plot.  }
\vspfigB
\label{fig3Qcor}\end{figure}  
%-------------------------------------------------------------------- 

   The correlation function for Helfand's moment of viscosity 
should be an even function of the shear rate. 
   Noting this point, in the small shear rate region of Fig. \ref{fig3Qcor} 
we give the fit of numerical data to the function 
$y=\alpha_{C0} + \alpha_{C2} x^{2}$ 
with the fitting parameter values 
$\alpha_{C0} = 46.749$ and 
$\alpha_{C2} = 1349.38$. 
%
%A20             = 46.749           +/- 10.38        (22.21%)
%A22             = 1349.38          +/- 14.6         (1.082%)
   The graph is nicely fitted by this quadratic function in 
the small shear rate region. 
   This point may be explained by noting 

\begin{eqnarray}
  \left\langle Q^{2}\right\rangle - \left\langle Q\right\rangle^{2} 
  &\approx& \sum_{j=1}^{N}\sum_{k=1}^{N}  
      \left\langle q_{jy} q_{ky} \tilde{p}_{jx} 
      \tilde{p}_{kx}\right\rangle 
  \nonumber \\
  &&+ \gamma^2 
      \sum_{j=1}^{N}\sum_{k=1}^{N}  
      \left(\left\langle q_{jy}^{2}q_{ky}^{2}\right\rangle
      - \left\langle q_{jy}^{2}\right\rangle 
      \left\langle q_{ky}^{2}\right\rangle\right)
  \nonumber \\
\label{HelfaCorri}\end{eqnarray}

\noindent using the thermal momentum component $\tilde{p}_{jx}$. 
   Here we assumed that the time-average of 
the linearly dependent terms for the thermal momentum  
can be neglected. 
   As the two terms $\sum_{j=1}^{N}\sum_{k=1}^{N}  
\langle q_{jy} q_{ky} \tilde{p}_{jx} \tilde{p}_{kx} \rangle$ 
and $\sum_{j=1}^{N}\sum_{k=1}^{N}  
\sum_{j=1}^{N}  \langle q_{jy}^{2}q_{ky}^{2}\rangle$ can be 
considered $\gamma$-independent, the correlation 
$\langle Q^{2} \rangle- \langle Q \rangle^{2}$  
can be fitted by a quadratic function of $\gamma$. 

   In the inset to Fig. \ref{fig3Qcor}, we give 
the shear rate dependence of the time-average 
of the correlation function for Helfand's moment of viscosity 
per unit volume $\mathcal{V}$ in a much wider region of shear rate 
in a linear-log scale. 
   (Note that we used a linear-linear scale in the main figure 
of Fig. \ref{fig3Qcor}.) 
   It should be noted that a rapid drop of the value of 
this correlation function occurs in the string phase region. 
   In the intermediate region, which is approximately 
the region $2.5<\gamma<8$ in Fig. \ref{fig3Qcor}, 
between the region fitted by the quadratic function of $\gamma$ 
and the string phase region, fluctuations in the value  
$\langle Q^{2} \rangle- \langle Q \rangle^{2}$ become 
much larger than in the other regions, and their values 
in Fig. \ref{fig3Qcor}
are less reliable.

%T---------------------------------------------------------------------

\subsection{Remarks in Connection with the Isokinetic Thermostat Dynamics and the Canonical Distribution Approach}
\label{remarksimul}

   Sllod dynamics with the isokinetic thermostat used 
in this section has been used very frequently to simulate 
shear flows. 
   It is supposed to reproduce the value of shear stress 
predicted by a canonical distribution approach \cite{Mor88}, and 
succeeded even to reproduce some real 
experimental values \cite{Mar01a}.  
   However one must notice that 
strictly speaking the time-average from Sllod dynamics with 
the isokinetic thermostat does not always reproduce 
the ensemble average for the non-equilibrium canonical 
distribution used in this paper,  
%   it is not exactly proper to 
%reproduce the canonical distribution itself, 
even in the equilibrium state where $\gamma=0$ 
after taking the thermodynamic limit $N\rightarrow\infty$ 
(and $\rho=const$). 
   Now we discuss a couple of examples illustrating these
   ensemble differences. 
   
   First, in the numerical simulations used in this section 
the sum of the thermal momentum $\tilde{\mathbf{p}}_{j}
\equiv(\tilde{p}_{jx},\tilde{p}_{jy}) 
\equiv\mathbf{p}_{j}-m\mathbf{V}(\mathbf{q}_{j})$ 
over the particle number $j$ in each direction is 
zero at all time, meaning that 
there is a constraint on the values of the thermal momenta, 
that is $\sum_{j=1}^{N}\tilde{p}_{jx}=0$. 
   On the other hand, in the canonical distribution 
approach all components of momenta can be 
treated as independent variables. 
   This difference, for example, causes the different averaged 
values for $\sum_{j=1}^{N}\sum_{k=1}^{N} 
\langle \tilde{p}_{jx}\tilde{p}_{kx}\rangle$ 
and $\sum_{j=1}^{N}\sum_{k=1}^{N} 
\overline{\tilde{p}_{jx}\tilde{p}_{kx}}$.  
   Actually the value of $\sum_{j=1}^{N}\sum_{k=1}^{N} 
\langle \tilde{p}_{jx}\tilde{p}_{kx}\rangle=
\langle (\sum_{j=1}^{N}\tilde{p}_{jx})(\sum_{k=1}^{N} 
\tilde{p}_{kx})\rangle$ 
is zero as each bracketed sum is individually zero. 
%in this section because it is the time-average of the square 
%of $x$-component of total thermal momentum,  
The value of $\sum_{j=1}^{N}\sum_{k=1}^{N} 
\overline{\tilde{p}_{jx}\tilde{p}_{kx}}$ 
however, is given by $mNT$ in the canonical distribution 
because of $\overline{\tilde{p}_{jx}\tilde{p}_{kx}} 
= mT \delta_{jk}$. 

   Second, the isokinetic thermostat used in 
the simulations of this section keeps the kinetic energy 
constant so that the distribution function for the 
kinetic energy is given by a delta function. 
   This is different from 
the distribution of kinetic energy derived from 
the canonical distribution, because there is 
always a non-zero 
fluctuation of the kinetic energy around its mean value 
in the canonical distribution. 
   Refs. \cite{Eva83b,Eva84} tried to modify the canonical 
distribution to give consistency with the isokinetic 
thermostat, but it is not obvious that we can justify 
the shear flow thermodynamics based on such a 
modified canonical distribution.

%---------------------------------------------------------------------
\begin{figure}[!htb]
\vspfigA
\includegraphics[width=\widthfigA]{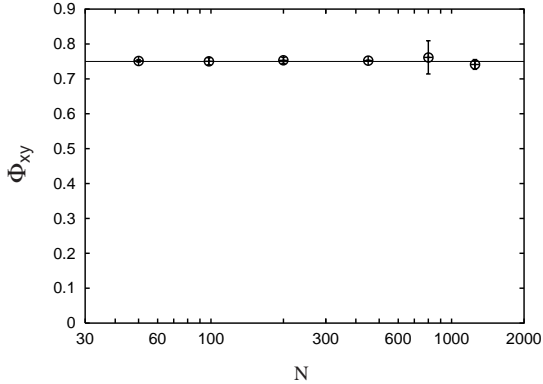}
\caption{
      Time-averaged quantity $\Phi_{xy}\equiv|\tilde{\Psi}_{xy}
   - \bar{\Psi}_{xy}|/\bar{\Psi}_{xy}$ 
   as a function of particle number $N$ for Sllod dynamics with 
   an isokinetic thermostat in a square at equilibrium 
   $\gamma=0$ with the particle density 
   $\rho=0.8$ as a log-linear plot.   
      Here $\tilde{\Psi}_{xy}$ and $\bar{\Psi}_{xy} $ 
   are defined by $\tilde{\Psi}_{xy} \equiv\sum_{j=1}^{N}\sum_{k=1}^{N}  
   \langle q_{jy} q_{ky} \tilde{p}_{jx} \tilde{p}_{kx}
   \rangle |_{\gamma=0}$
   and $\bar{\Psi}_{xy} \equiv mT \sum_{j=1}^{N} 
   \langle q_{jy}^{2}\rangle|_{\gamma=0}$, respectively. 
%   give the same value in ensemble average of the canonical distribution. 
      The length of error bars in this figure is given 
   by $2|\Phi_{xy}-\Phi_{yx}|$. 
      The solid line is the value $3/4$, which is explained in the text.    
   }
\vspfigB
\label{fig4QPerr}\end{figure}  
%-------------------------------------------------------------------- 

   As a concrete example of these ensemble  
differences, let's consider 
the first term $\tilde{\Psi}_{xy}
\equiv\sum_{j=1}^{N}\sum_{k=1}^{N}  
\langle q_{jy} q_{ky} \tilde{p}_{jx} \tilde{p}_{kx}
\rangle |_{\gamma=0}$
appearing on the right-hand side of Eq. (\ref{HelfaCorri}) 
at equilibrium $\gamma=0$. Assuming that the variables 
$q_{jy}$ and  $\tilde{p}_{jx}$ are independent, and that 
$\langle q_{jy} q_{ky} \rangle$ only depends upon whether 
$j=k$ or $j \neq k$, then 

\begin{eqnarray}
   &&\hspace{-0.5cm} \sum_{j=1}^{N}\sum_{k=1}^{N}  
      \left\langle q_{jy} q_{ky} \tilde{p}_{jx} \tilde{p}_{kx} \right\rangle
      \nonumber \\
   &&\hspace{-0.5cm} \approx
      \left\langle q_{1y}^{2} \right\rangle 
      \sum_{j=1}^{N} \left\langle \tilde{p}_{jx}^{2}\right\rangle
      +\left\langle q_{1y} q_{2y}\right\rangle 
         \sum_{j=1}^{N} \sum_{k=1 (k\neq j)}^{N} 
      \left\langle\tilde{p}_{jx}\tilde{p}_{kx}\right\rangle
\label{EnsDiff}\end{eqnarray}

\noindent If $q_{1y}$ is uniformly distributed between $0$ and $L$ then 
$\langle q_{1y}^{2}\rangle \approx L^{2}/3$ and 
$\langle q_{1y} q_{2y} \rangle \approx L^{2}/4$. 
%If this time-averaged quantity in Sllod dynamics 
%could be replaced by 
%its ensemble average in the canonical distribution, then 
%the average value $\bar{\Psi}_{xy}$ should 
%be equal to $\tilde{\Psi}_{xy} 
%\equiv mT \sum_{j=1}^{N} \langle q_{jy}^{2}
%\rangle |_{\gamma=0}$. 
   The difference between 
$\bar{\Psi}_{xy}$ and $\tilde{\Psi}_{xy}$ 
can be calculated as
$\overline{\tilde{p}_{jx}\tilde{p}_{kx}} =0$ in $j\neq k$ 
for the canonical average, whereas for the time average
$\sum_{j\neq k}\langle\tilde{p}_{jx}\tilde{p}_{kx}\rangle 
=- \sum_{j=1}^{N}\langle\tilde{p}_{jx}^{2}\rangle$. 
It follows that $\bar{\Psi}_{xy}\approx mNTL^{2}/3$ and 
$\tilde{\Psi}_{xy} \approx mNTL^{2}/12$, so $\Phi_{xy} 
\equiv |\tilde{\Psi}_{xy}
- \bar{\Psi}_{xy}|/\bar{\Psi}_{xy} \approx 3/4$. 
%in the numerical model used in this section. 
   Figure \ref{fig4QPerr} is the graph of the normalized 
difference $\Phi_{xy}$ 
%$= |\bar{\Psi}_{xy}- \tilde{\Psi}_{xy}|/\tilde{\Psi}_{xy}$ 
as a function of system size $N$ for square systems
at fixed density $\rho=0.8$ in numerical simulations. 
%   This quantity should be zero, if the time-average 
%in this simulation is equivalent to the ensemble average 
%of the canonical distribution. 
   The length of error bars in this figure is given 
by $2|\Phi_{xy}-\Phi_{yx}|$ which must be zero in the square cases. 
   Figure \ref{fig4QPerr} suggests that $\Phi_{xy}$ is in excellent
agreement with the value of $3/4$ given above.  
%up to $N=1250$, and it is not convincing    
%that $\Phi_{xy}$ will go to zero in the thermodynamic limit 
%$N\rightarrow\infty$ (with $\rho=const$). 

%\subsection{Helfand Moment of Viscosity and its Correlation}
%\subsection{Pressure-Stress Tensor}

%%%%%%%%%%%%%%%%%%%%%%%%%%%%%%%%%%%%%%%%%%%%%%%%%%%%%%%%%%%%%%%%%%%%%%

\section{Conclusion and Remarks}
\label{ConclusionRemarks}

   In this paper we have discussed a canonical distribution 
approach to non-equilibrium steady flows 
for the purpose of constructing a steady state thermodynamics
from solid statistical mechanical foundations. 
   Using the Lagrangian technique of classical mechanics 
we introduced the energy in the moving frame 
by separating the velocity of the global steady flow. 
   A canonical distribution based on this internal energy 
was introduced. 
   As one application of this distribution, 
we showed that the well known thermodynamics of rotating 
systems can be derived from this canonical distribution. 
   Our special concern in this canonical distribution approach 
was steady shear flows and their thermodynamics. 
   Evans and Hanley proposed a 
first law of thermodynamics of the form 
$d\mathcal{E} = T d\mathcal{S} - \mathcal{Q}d\gamma$ relating 
energy $\mathcal{E}$, temperature $T$, entropy $\mathcal{S}$ 
and shear rate $\gamma$. 
   Here we derived this shear flow thermodynamics based on our canonical 
distribution approach, 
and showed that the quantity $\mathcal{Q}$ is 
given by the average of Helfand's moment of viscosity, 
the temperature $T$ 
is the kinetic temperature derived from the thermal kinetic energy, 
and $\mathcal{E}$ can be interpreted as an internal energy. 
   The roles of the work required to sustain the shear flow 
and the heat removed to compensate it (the house-keeping heat) 
was emphasized in the justification of the linear response 
formula for viscosity, which is derived 
from our shear flow canonical distribution approach. 
   We introduced a non-equilibrium entropy, and showed 
that it increases in time and the house-keeping heat based on this 
entropy has the same magnitude as the power needed to sustain the 
steady flow. 
   This discussion led to the non-negativity of average of 
$-\gamma P_{xy}$ with the shear stress $P_{xy}$, meaning that 
the power needed to sustain the shear flow and 
the house-keeping heat is always non-negative. 
   We discussed the thermodynamic stability condition 
for the shear flows, one of which is equivalent to 
the positivity of the correlation function of Helfand's moment 
of viscosity.
   Our results were investigated in numerical simulations of 
two-dimensional many-particle systems with soft-core interactions, 
whose dynamics is determined by the Sllod equations with an 
isokinetic thermostat. 

   To construct the canonical distribution for shear flow, 
we used the analogy of shear flows and rotating systems. 
   These two systems are steady flows whose magnitude is proportional 
to a component of position vector: the distance from the rotating axis 
in the rotating system, and the position component orthogonal to 
the flow in the shear system. 
   Both systems also have clear parameters to characterize 
their currents: the angular velocity in the rotating system 
and the shear rate in the shear flow. 
   On the other hand, we also emphasized some differences 
between these two systems. 
   The biggest difference may be that the total angular momentum 
in the canonical distribution of the rotating system 
is time-independent, whereas Helfand's moment of viscosity 
appearing the canonical distribution of the shear flow 
is not constant. 
   This led to the necessity to consider the work needed to sustain 
the steady flow and the house-keeping heat in the shear flow system, 
and plays an essential role in the derivation 
of the response formula for viscosity.    

   One may easily notice that the canonical distribution 
approach discussed in this paper can 
be generalized to more general steady flows than 
the rotating system and the shear flow system. 
   One of the restrictions in our canonical distribution 
approach is that we have to know the global velocity 
distribution $\mathbf{V}$ \emph{a priori}. 
   In this sense this approach is not appropriate to determine 
the global velocity distribution under some external 
constraints, etc. 
   It is also crucial that we know a priori 
an external parameter that specifies the amount of the 
global flow, like the angular velocity or the 
the shear rate. 
  This parameter is treated as a thermodynamic 
quantity in the expression for the first law of thermodynamics. 

%   The Evans-Hanley's thermodynamics of shear flow adopts 
%the form close to the equilibrium thermodynamics, 
%but we should notice that there are some significant 
%difference between the two. 
%   Foe example, the energy $\overline{H^{(ine)}}$ 
%in the inertial frame $\mathcal{F}^{(ine)}$ is not 
%an extensive quantity in the orthogonal direction 
%of the shear velocity. 

%---------------------------------------------------------------------
\begin{figure}[!htb]
\vspfigA
\includegraphics[width=\widthfigB]{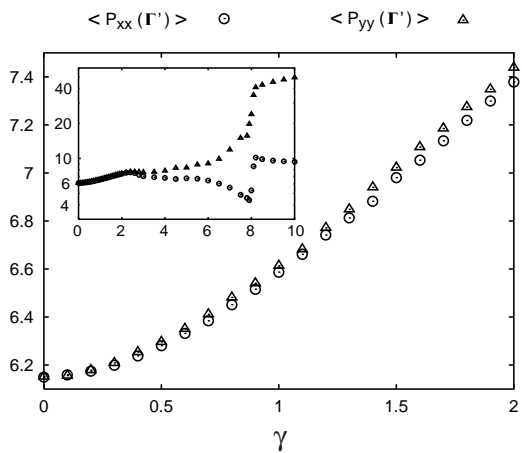}
\caption{
      Time-averages of the diagonal components 
   $\langle P_{xx}(\bfGamma')\rangle$ (circles) 
   and $\langle P_{yy}(\bfGamma')\rangle$ (triangles) of 
   the pressure tensor with the 
   thermal phase space vector $\bfGamma'$ as functions of 
   shear rate $\gamma$ in Sllod dynamics with 
   an isokinetic thermostat in a linear-linear plot. 
      The inset: the same graph except for including 
   a wider region of shear rate and in a linear-log plot. 
   }
\vspfigB
\label{fig5PxxPyy}\end{figure}  
%-------------------------------------------------------------------- 

   An important future problem in  
shear flow thermodynamics based on the canonical distribution 
approach is to discuss the pressure in this framework. 
   Refs. \cite{Eva80a,Eva80b,Han82,Eva83} introduced 
the pressure $\mathcal{P}$ 
simply by adding the term $-\mathcal{P}d\mathcal{V}$ 
on the right-hand side of Eq. (\ref{EvansHanley}). 
   For this term it was conjectured that the pressure 
$\mathcal{P}$ would be equal to the minimum eigenvalue 
of the pressure tensor \cite{Eva89}. 
   However one should notice that non-equilibrium systems 
such as the shear flow system are not generally isotropic, 
so that the pressure defined by $-\partial \mathcal{E} 
/ \partial \mathcal{V}$ may  
depend on which direction we change the volume 
$\mathcal{V}$. 
   Actually, as shown in Fig. \ref{fig5PxxPyy}, 
the numerical simulations using Sllod dynamics in 
Sec. \ref{numeritest} show that the time averages of 
$P_{xx}(\bfGamma')$ and $P_{yy}(\bfGamma')$ 
are different from each other at non-zero shear rate.
   (Here $\bfGamma'$ is the "thermal phase space vector" 
introduced as the vector in which 
the momentum $\mathbf{p}_{j}$ in the phase space vector 
is replaced by the thermal momentum 
$\tilde{\mathbf{p}}_{j}\equiv \mathbf{p}_{j} 
- m \mathbf{V}(\mathbf{q}_{j})$.) 
   Noting that usually the pressure 
is calculated by the arithmetic average of these time-averages 
(or ensemble averages),  
%$P_{xx}(\bfGamma')$ and $P_{yy}(\bfGamma')$ 
(See Ref. \cite{Eva90a}, also Ref. \cite{Zub74} for 
its justification using the microcanonical distribution.), 
this suggests that if the pressures in the $x$  
and the $y$-directions are given by averages of 
$P_{xx}(\bfGamma')$ and $P_{yy}(\bfGamma')$, respectively, 
then the pressure is direction-dependent 
in shear flow systems. 
   The quantity $\langle P_{xx}(\bfGamma')\rangle 
- \langle P_{yy}(\bfGamma')\rangle$ is called 
the "normal stress" and a non-zero value is 
one of the important properties of visco-elastic 
fluids \cite{Tsh89,Eva90a,Que81}. 
   Therefore it is important to understand 
whether such a property is compatible with 
the thermodynamical framework discussed in this paper, 
in other words, to discuss the first law of thermodynamics 
in which the averages $\overline{P_{xx}(\bfGamma')}$ and 
$\overline{P_{yy}(\bfGamma')}$ are 
included as the $x$ and $y$-components 
of the pressure, respectively. 
   It may be noted that a similar question can be asked 
for rotating systems. 
%   These points will be discussed in the separate paper. 
   We leave discussion of these points for the future. 

%The local distribution is not the global variable in our approach
%---> It make our approach consistent with the Evans-Hanley's 
%     thermodynamics, which introduce an external 
%     parameter to manipulate shear flow as one of the thermodynamical 
%     quantities, such as the external magnetic field of 
%     thermodynamics of magnetism and the angular velocity 
%     of the thermodynamics of the rotating system. 

   As mentioned in Sec. \ref{twoaverage}, 
the thermodynamic relations,  
(\ref{SearFirstLaw4}) and (\ref{ShearStabi3d}) derived 
in this paper,  are relations for the ensemble average 
(\ref{Avera}) under the canonical distribution $f(\bfGamma)$. 
   On the other hand the numerical calculations discussed 
in Sec. \ref{numeritest} give the average 
(\ref{ShearAvera}) under the distribution $\tilde{f}(\bfGamma,\infty)$.  
   Although these two averages are related by Eq. (\ref{TwoAvera}), 
it is still an open question 
to calculate the canonical average (\ref{Avera}), 
required for the thermodynamic relations, from the dynamical evolved 
canonical average (\ref{ShearAvera}) 
%to check quantitatively 
%the thermodynamic relation using the average (\ref{Avera}) 
%from the average (\ref{ShearAvera}) given 
in numerical calculations. 

   Originally, Evans and Hanley introduced 
their shear flow thermodynamics 
to discuss non-analytical properties of the pressure, viscosity 
and the internal energy as functions of the shear rate. 
   Such non-analytical properties are predicted by 
mode-coupling theory \cite{Kaw73,Pom75}, 
and are supported by some numerical 
calculations \cite{Eva90a,Eva80c,Eva81}. 
   However, recently some numerical works suggest that 
the shear rate dependence of the pressure is 
rather analytic near equilibrium, except at the triple 
point \cite{Mar01a,Ge01}. 
   Moreover, even at the triple point the non-analytic 
dependence of the pressure is not completely 
convincing \cite{Tra95}.
   It may also be noted that some theories, that predict 
an analytic dependence of the pressure and the viscosity 
with respect to the shear rate, have been proposed 
\cite{Jou01,Tra95,Que81}. 
   In this sense it is still an interesting problem 
to discuss shear rate dependences of the pressure, 
the viscosity and so on using shear flow thermodynamics.

%---------------------------------------------------------------------
\begin{figure}[!htb]
\vspfigA
\includegraphics[width=\widthfigB]{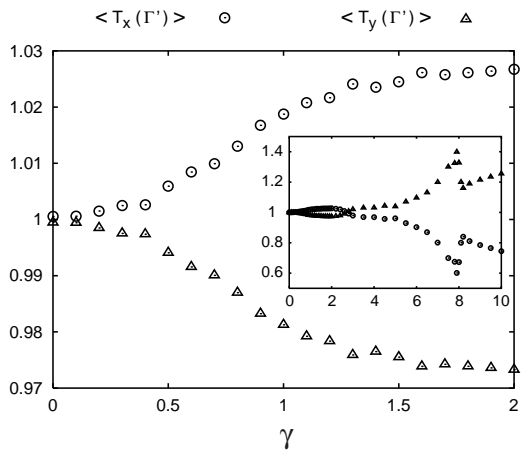}
\caption{
      $x$-component $\langle T_{x}(\bfGamma') \rangle$ 
   (circles) and $y$-component 
   $\langle T_{y}(\bfGamma') \rangle$ 
   (triangles) of the kinetic temperature with the 
   thermal phase space vector $\bfGamma'$
   as functions of shear rate $\gamma$  in the Sllod dynamics with 
   the isokinetic thermostat. 
      The inset: the same graph except for including a wider range 
   of shear rate. 
   }
\vspfigB
\label{fig6TxTy}\end{figure}  
%-------------------------------------------------------------------- 

   There are also questions about the numerical simulations 
of shear flow themselves, from the point of view of the 
canonical distribution approach. 
   Some such problems were already mentioned 
in Sec. \ref{remarksimul}. 
   As another potential problem, we mention the direction dependence 
of the thermal kinetic energy. 
   To discuss this point we introduce the quantities 
$T_{x}(\bfGamma')$ and $T_{y}(\bfGamma')$ as 
%\begin{eqnarray}
$
   T_{k}(\bfGamma') \equiv 
   (2/N)\sum_{j=1}^{N} \tilde{p}_{jk}^{2}/(2m)
$
%\label{TempeTxTy}\end{eqnarray}
%\noindent 
with the thermal momentum component 
$\tilde{p}_{jk}\equiv 
p_{jk} - m V_{k}(\textbf{q}_{j})$, where $V_{k}$ 
is the $k$-component of the 
global current density $\mathbf{V}$. 
   The arithmetic average of ensemble averages of 
$\{T_{k}(\bfGamma')\}_{k}$ over the component $k$ 
gives the kinetic temperature, so we may interpret 
the quantity $T_{k}(\bfGamma')$ as the observable for 
the "$k$-component of the temperature". 
   The canonical distribution approach discussed in this paper 
claims that the ensemble average $\overline{T_{k}(\bfGamma')}$ 
of the quantity $T_{k}(\bfGamma')$ is $k$-independent, 
in other words 
the kinetic temperature is direction-independent, although 
we should note a difference in the two averages 
$\overline{T_{k}(\bfGamma')}$ and 
$\langle T_{k}(\bfGamma') \rangle_{\infty}$. 
   Figure \ref{fig6TxTy} shows the graphs of 
$\langle T_{x}(\bfGamma') \rangle$ 
and $\langle T_{y}(\bfGamma') \rangle$ 
as functions of shear rate $\gamma$ from numerical 
simulations using the Sllod dynamics with an isokinetic thermostat, 
used in Sec. \ref{numeritest}. 
   This figure shows that the kinetic temperature is  
direction-dependent at least in large shear rate cases. 
   As a topic related  to this point, it may be noted that in 
the isokinetic thermostat the heat is removed from any component 
of kinetic energy of any particle uniformly. 
   This gives great simplification in the formula and 
numerical calculations and keeps a similar dynamical structure 
to Hamiltonian dynamics leading to the numerical observation of 
the conjugate pairing rule for the Lyapunov spectrum \cite{Mor01}, 
but its physical justification as a mechanical thermostat is not 
completely convincing. 
   For example, one may use other types of thermostats 
in which the heat is removed from the particles near the walls 
or from the kinetic energy component orthogonal to the walls 
\cite{Tro84,Pos89}. 
%considered thermostats working 
%only for particles to hit walls. 
   These different thermostats might give, for example, 
different values of $\langle T_{x}(\bfGamma') \rangle$ 
and $\langle T_{y}(\bfGamma') \rangle$ from the 
isokinetic thermostat. 
   To check the shear flow thermodynamics 
for such types of thermostat remains an open problem.

%%%%%%%%%%%%%%%%%%%%%%%%%%%%%%%%%%%%%%%%%%%%%%%%%%%%%%%%%%%%%%%%%%%%%%

\section*{Acknowledgements}

   We wish to thank P. Daivis for valuable discussions 
on the content of this paper. 
   We are grateful for the financial support for this work 
from the Australian Research Council. 
   One of the authors (T.T.) also appreciates the financial support 
by the Japan Society for the Promotion Science. 
%and a hospitality of University of New South Wales. 

%%%%%%%%%%%%%%%%%%%%%%%%%%%%%%%%%%%%%%%%%%%%%%%%%%%%%%%%%%%%%%%%%%%%%%

\appendix
\section{Response Formula for the Viscosity from the Canonical Distribution Approach}
\label{GreKubVisco}

   In this appendix we give a derivation of the linear response 
formula (\ref{GreenKuboVisco}) for viscosity 
from the definition (\ref{visco}), 
as well as a derivation of Eq. (\ref{zerothvisco}). 
   We also discuss  
the two kinds of nonlinear response formulas for 
$\left\langle P_{xy}(\bfGamma) \right\rangle_{\infty}$ 
with respect to the shear rate $\gamma$, 
one of which is a simple generalization of 
the formula (\ref{GreenKuboVisco}).

   First we note that the partition function $\Xi$ 
can be rewritten as 

\begin{eqnarray}
   \Xi&\equiv&\int d\bfGamma 
     \exp\left\{-\beta H^{(mov)}(\bfGamma)\right\}
     \nonumber \\
   &=&\int d\bfGamma 
     \exp \left\{-i\hat{L}^{(ine)} (t-t_{0})
     \right\}
     \nonumber \\
   &&\eqspaB \times
     \exp\left\{-\beta H^{(mov)}(\bfGamma)\right\}
     \nonumber \\
   &=&\int d\bfGamma 
     \exp\left\{-\beta H^{(mov)}(\bfGamma)\right\}
     \nonumber \\
   &&\eqspaB \times
     \exp\left\{ -\beta
     \gamma \mathcal{V} 
     \int_{t_{0}}^{t} ds \tilde{P}_{xy}(\bfGamma,-s+2t_{0})
     \right\} \;\;\;
\label{Parti}\end{eqnarray}

\noindent where we used the relations
$\exp\{-i\hat{L}^{(ine)} (t-t_{0}) \} 1 = 1$, 
$(i \hat{L}^{(ine)})^{\dagger} = 
- i \hat{L}^{(ine)}$ ($\dagger$ meaning to take its Hermitian conjugate)
and a similar derivation to that in Eq. (\ref{GenerShearCanon2}). 
    Equation (\ref{Parti}) means that both the distributions 
$f(\bfGamma)$ and $\tilde{f}(\bfGamma,t)$ are normalized 
with the same partition function $\Xi$. 
   The partition function $\Xi$ given by Eq. (\ref{Parti}) 
must be time-independent, so that we obtain 

\begin{eqnarray}
   0 &=& \frac{\partial \Xi}{\partial t}
     \nonumber \\
   &=&-\beta
     \gamma \mathcal{V} 
     \int d\bfGamma 
     \tilde{P}_{xy}(\bfGamma,-t+2t_{0})
     \nonumber \\
   &&\eqspaB \times
     \exp\left\{-\beta H^{(mov)}(\bfGamma)\right\}
     \nonumber \\
   &&\eqspaB \times
     \exp\left\{ -\beta
     \gamma \mathcal{V} 
     \int_{t_{0}}^{t} ds \tilde{P}_{xy}(\bfGamma,-s+2t_{0})
     \right\} 
     \nonumber \\
   &=&-\beta
     \gamma \mathcal{V} \Xi 
     \int d\bfGamma 
     \tilde{P}_{xy}(\bfGamma,-t+2t_{0})
     \nonumber \\
   &&\eqspaB \times 
     \exp \left\{-i\hat{L}^{(ine)} (t-t_{0})
     \right\} f(\bfGamma)
     \nonumber \\
   &=&-\beta
     \gamma \mathcal{V} \Xi 
     \int d\bfGamma 
     \exp \left\{-i\hat{L}^{(ine)} (t-t_{0})
     \right\} 
     \nonumber \\
   &&\eqspaB \times
     P_{xy}(\bfGamma) 
     f(\bfGamma)
     \nonumber \\
%   &=&-\beta
%     \gamma \mathcal{V} \Xi 
%     \int d\bfGamma 
%     P_{xy}(\bfGamma) 
%     f(\bfGamma) 
%     \nonumber \\
   &=&-\beta
     \gamma \mathcal{V}  \Xi 
     \overline{P_{xy}(\bfGamma)}   
\label{Parti2}\end{eqnarray} 

\noindent noting the definition of the 
average (\ref{Avera}). 
   From Eq. (\ref{Parti2}) we obtain Eq. (\ref{zerothvisco}), 
implying that the viscosity calculated 
from the canonical distribution $f(\bfGamma)$ is zero. 

%\begin{eqnarray}
%     \overline{P_{xy}(\bfGamma)} = 0 
%\label{Parti3}\end{eqnarray} 

   On the other hand, using the average (\ref{ShearAvera}) 
by the distribution $\tilde{f}(\bfGamma,t)$ given by 
Eq. (\ref{GenerShearCanon2}) we have 

\begin{widetext}
\begin{eqnarray}
   \left\langle P_{xy}(\bfGamma) \right\rangle_{\infty} 
   &=& \lim_{t\rightarrow \infty} \int d\bfGamma 
      P_{xy}(\bfGamma) 
      f(\bfGamma) \exp\left\{ -\beta
      \gamma \mathcal{V} 
      \int_{t_{0}}^{t} ds \tilde{P}_{xy}(\bfGamma,-s+2t_{0})
   \right\} 
   \nonumber \\ &=& 
      \int d\bfGamma 
      P_{xy}(\bfGamma) 
      f(\bfGamma) 
%      \nonumber \\
%   && \hspace{1cm} 
      -\beta
      \gamma \mathcal{V}  
      \int_{t_{0}}^{\infty} ds \int d\bfGamma 
      P_{xy}(\bfGamma) 
      f(\bfGamma) 
      \tilde{P}_{xy}(\bfGamma,-s+2t_{0})
%      \nonumber \\
%   && \hspace{1cm} 
      +  \mathcal{O}(\gamma^{2}) 
      \nonumber \\
   &=& \overline{P_{xy}(\bfGamma)} 
      -\beta
      \gamma \mathcal{V}
      \int_{t_{0}}^{\infty} ds \int d\bfGamma 
      P_{xy}(\bfGamma) f^{(eq)}(\bfGamma) 
      \tilde{P}_{xy}(\bfGamma,-s+2t_{0}) 
%      \nonumber \\
%   && \hspace{1cm} 
      +  \mathcal{O}(\gamma^{2}) 
      \nonumber \\
   &=& -\beta
      \gamma \mathcal{V}
      \int_{t_{0}}^{\infty} ds \int d\bfGamma 
      P_{xy}(\bfGamma) f^{(eq)}(\bfGamma) 
      \exp\left\{ - i \hat{L}^{(ine)} (s-t_{0})
      \right\}
      P_{xy}(\bfGamma) 
%      \nonumber \\
%   && \hspace{1cm} 
      +  \mathcal{O}(\gamma^{2}) 
      \nonumber \\
   &=& -\beta
      \gamma \mathcal{V}
      \int_{t_{0}}^{\infty} ds \int d\bfGamma 
      \left[
       \exp\left\{ i \hat{L}^{(ine)} (s-t_{0})
      \right\} P_{xy}(\bfGamma) f^{(eq)}(\bfGamma) 
     \right]
      P_{xy}(\bfGamma) 
%      \nonumber \\
%   && \hspace{1cm}
      +  \mathcal{O}(\gamma^{2}) 
      \nonumber \\
   &=& -\beta
      \gamma \mathcal{V}
      \int_{t_{0}}^{\infty} ds \left\langle
      \tilde{P}_{xy}(\bfGamma,s) 
      P_{xy}(\bfGamma)  \right\rangle^{(eq)} 
      +  \mathcal{O}(\gamma^{2}) 
\label{0thviscoDerB1}\end{eqnarray} 
\end{widetext}

\noindent with the notation 
$f^{(eq)}(\bfGamma) \equiv \lim_{\gamma\rightarrow 0} 
f(\bfGamma)$, where we used Eq. (\ref{zerothvisco}), 
the relation $(i \hat{L}^{(ine)})^{\dagger} = 
- i \hat{L}^{(ine)}$, and 
$\exp\{ i \hat{L}^{(ine)} (s-t_{0}) \} f^{(eq)}(\bfGamma) 
= f^{(eq)}(\bfGamma)$. 
   Equation (\ref{0thviscoDerB1}) leads to 
the linear response formula (\ref{GreenKuboVisco}) 
for viscosity.

   Next, using Eq. (\ref{GenerShearCanon2}) we have 

\begin{eqnarray}
   \frac{\partial \tilde{f}(\bfGamma,t)}{\partial t} 
   = -
%      \left[
      \beta
      \gamma \mathcal{V} \tilde{P}_{xy}(\bfGamma,-t+2t_{0}) 
%      \right]
      \tilde{f}(\bfGamma,t)
\label{DerivTildF}\end{eqnarray}   
   
\noindent The solution of the time-differential equation 
(\ref{DerivTildF}) of the function 
$ \tilde{f}(\bfGamma,t)$ with the initial condition 

\begin{eqnarray}
   \tilde{f}(\bfGamma,t_{0}) = f(\bfGamma)
\label{InitialF}\end{eqnarray} 

\noindent is represented as 

\begin{widetext}
\begin{eqnarray}
   \tilde{f}(\bfGamma,t) &=& f(\bfGamma) 
   + \sum_{n=1}^{\infty} \left(-\beta
      \gamma \mathcal{V}\right)^{n}
   \int_{t_{0}}^{t} ds_{1} \int_{t_{0}}^{s_{1}} ds_{2} 
   \int_{t_{0}}^{s_{2}} ds_{3} 
   \cdots \int_{t_{0}}^{s_{n-1}} ds_{n} 
   \nonumber \\ 
   && \hspace{2cm} \times \tilde{P}_{xy}(\bfGamma,-s_{1}+2t_{0}) 
   \tilde{P}_{xy}(\bfGamma,-s_{2}+2t_{0}) 
   \cdots \tilde{P}_{xy}(\bfGamma,-s_{n}+2t_{0})
   f(\bfGamma) .
\label{TildFExpa}\end{eqnarray} 

\noindent From Eqs. (\ref{zerothvisco}) and (\ref{TildFExpa}) we derive 

\begin{eqnarray}
%   &&
   \left\langle P_{xy}(\bfGamma) \right\rangle_{\infty} 
%   \nonumber \\ 
%   &&\hspace{0.5cm} 
   &=& \sum_{n=1}^{\infty} \left(-\beta
      \gamma \mathcal{V}\right)^{n}
   \int_{t_{0}}^{\infty} ds_{1} \int_{t_{0}}^{s_{1}} ds_{2} 
   \int_{t_{0}}^{s_{2}} ds_{3} 
   \cdots \int_{t_{0}}^{s_{n-1}} ds_{n} 
   \nonumber \\ 
   && \hspace{2cm} \times \overline{ P_{xy}(\bfGamma)
   \tilde{P}_{xy}(\bfGamma,-s_{1}+2t_{0}) 
   \tilde{P}_{xy}(\bfGamma,-s_{2}+2t_{0}) 
   \cdots \tilde{P}_{xy}(\bfGamma,-s_{n}+2t_{0})}. 
\label{PxyExpan}\end{eqnarray} 
\end{widetext}

\noindent This expresses a nonlinear response formula 
for an average of the shear stress $P_{xy}(\bfGamma)$ with respect to 
the shear rate $\gamma$ in the form of its 
multiple time-correlation function. 
   The formula (\ref{GreenKuboVisco}) can be derived 
directly from Eq. (\ref{PxyExpan}), using  
%Eq. (\ref{zerothvisco}) and 
the relations $\overline{X(\Gamma)}|_{\gamma = 0} = 
\langle X(\Gamma)|_{\gamma = 0}\rangle^{(eq)}$ 
in any function $X(\Gamma)$ of $\Gamma$ 
and $(i \hat{L}^{(ine)})^{\dagger} = 
- i \hat{L}^{(ine)}$.
   It may be noted that the multi-time integral functions 
on the right-hand side of Eq. (\ref{PxyExpan}) 
can be $\gamma$-dependent because of the $\gamma$-dependence of the 
function $f(\bfGamma)$, so strictly speaking 
Eq. (\ref{PxyExpan}) is not an expansion formula for 
$\left\langle P_{xy}(\bfGamma) \right\rangle_{\infty}$ 
with respect to the shear rate $\gamma$. 

   It may be meaningful to show another type of 
nonlinear response formula 
for the average of the quantity $P_{xy}(\bfGamma)$ with respect to 
the shear rate $\gamma$, using a Green's function $\hat{G}$ defined by 

\begin{eqnarray}
   \hat{G} \equiv \lim_{\epsilon\rightarrow +0} 
   \left[\hat{L}^{(ine)}+i\epsilon\right]^{-1}. 
\label{GreenFunct}\end{eqnarray} 

\noindent For this purpose, first we note a formal identity 

\begin{eqnarray}
   && \lim_{t\rightarrow+\infty} 
      \exp\left\{i\hat{L}^{(ine)} t\right\} 
   \nonumber \\ 
   &&\eqspaA = \lim_{\epsilon\rightarrow+0} \epsilon 
      \int_{0}^{+\infty} dt 
      \exp\left\{-\epsilon t\right\} \cdot 
      \exp\left\{i\hat{L}^{(ine)} t\right\} 
   \nonumber \\ 
   &&\eqspaA = \lim_{\epsilon\rightarrow+0} \epsilon 
   \left[\epsilon-i\hat{L}^{(ine)}\right]^{-1}
   \nonumber \\ 
   &&\eqspaA = 1- \hat{G} \hat{L}^{(ine)} , 
\label{limitepsilon}\end{eqnarray} 

\noindent where we used Eqs. (\ref{GreenFunct}) and 
$\epsilon [\epsilon-i\hat{L}^{(ine)}]^{-1} 
=[\epsilon-i\hat{L}^{(ine)}]^{-1} 
[\epsilon-i\hat{L}^{(ine)}+i\hat{L}^{(ine)}] 
= 1- \hat{G} \hat{L}^{(ine)}$. 
   Equation (\ref{limitepsilon}) is an analogous technique 
to that used in quantum scattering theory \cite{Joa75}
in which the Hamiltonian operator instead of the Liouville operator 
$\hat{L}^{(ine)}$ is used. 
   Using  Eq. (\ref{limitepsilon}) we have 
   
\begin{eqnarray}
   &&\left\langle P_{xy}(\bfGamma) \right\rangle_{\infty} 
   \nonumber \\ 
   &&\eqspaA = \lim_{t-t_{0}\rightarrow\infty} 
   \int d \bfGamma f(\bfGamma) 
   \exp\left\{i\hat{L}^{(ine)} (t-t_{0})\right\} 
   P_{xy}(\bfGamma)
   \nonumber \\ 
   &&\eqspaA =
   \int d \bfGamma f(\bfGamma) 
   \left[1- \hat{G} \hat{L}^{(ine)}\right] 
   P_{xy}(\bfGamma)
   \nonumber \\ 
   &&\eqspaA =
   \sum_{n=0}^{\infty} \frac{(\beta\gamma)^{n}}{n!} 
   \left\langle \left[Q(\bfGamma)\right]^{n} 
   \left[1- \hat{G} \hat{L}^{(ine)}\right] 
   P_{xy}(\bfGamma) \right\rangle^{(eq)} .
   \nonumber \\ 
   &&\eqspaA 
\label{PxyExpan2}\end{eqnarray} 

\noindent This is the formula which we wanted to derive. 
   It may be noted that the quantity 
$\langle [Q(\bfGamma)]^{n} [1- \hat{G} \hat{L}^{(ine)}] 
P_{xy}(\bfGamma) \rangle^{(eq)} $ appearing 
on the right-hand side 
of Eq. (\ref{PxyExpan2}) is $\gamma$-independent, so 
Eq. (\ref{PxyExpan2}) can be regarded as a real expansion 
of $\left\langle P_{xy}(\bfGamma) \right\rangle_{\infty}$ 
with respect to the shear rate $\gamma$, different 
from the formula (\ref{PxyExpan}). 
   Another merit of the formula (\ref{PxyExpan2}) 
is that we do not have to calculate a time-integral 
in the interval $[0,\infty]$, which is 
required in the formula (\ref{PxyExpan}). 
   As a special case of the formula (\ref{PxyExpan2}), 
using Eq. (\ref{visco}) and the fact that zero-th 
order of the quantity 
$\left\langle P_{xy}(\bfGamma) \right\rangle_{\infty}$ 
must be zero, we obtain 

\begin{eqnarray}
   \left\langle 
   \left[1- \hat{G} \hat{L}^{(ine)}\right] 
   P_{xy}(\bfGamma) \right\rangle^{(eq)} = 0, 
\label{specialcase1}\end{eqnarray} 
\begin{eqnarray}
   \eta =  - \beta\left\langle Q(\bfGamma)
   \left[1- \hat{G} \hat{L}^{(ine)}\right] 
   P_{xy}(\bfGamma) \right\rangle^{(eq)}. 
\label{specialcase2}\end{eqnarray} 
   
\noindent Eq. (\ref{specialcase2}) is another type 
of the linear response formula for the viscosity.

%%%%%%%%%%%%%%%%%%%%%%%%%%%%%%%%%%%%%%%%%%%%%%%%%%%%%%%%%%%%%%%%%%%%%%

\section{Canonical Distribution Approach using the Sllod Equation}
\label{SllodCanonical}

   In this appendix we discuss briefly the 
non-equilibrium canonical distribution 
approach using the Sllod equation for shear flows, 
and the relation 
of this approach to that discussed in the text of this paper. 

   The Sllod equation of shear flows is introduced as 
the equations for $\cald$-dimensional 
vectors $\mathbf{q}'_{j}(t)$ and $\mathbf{p}'_{j}(t)$ 
\cite{Eva90a}

\begin{eqnarray}
   \frac{d\mathbf{q}'_{j}(t)}{dt} = \frac{1}{m} \mathbf{p}'_{j}(t) 
   + \gamma \Theta \mathbf{q}'_{j}(t)
\label{SllodThermEquat1}\end{eqnarray}
\begin{eqnarray}
   \frac{d\mathbf{p}'_{j}(t)}{dt} = - \frac{\partial U(\mathbf{q}'(t))}
   {\partial\mathbf{q}'_{j}(t)} 
   - \gamma \Theta \mathbf{p}'_{j}(t)
\label{SllodThermEquat2}\end{eqnarray}

\noindent where $U(\mathbf{q}'(t))$ is the potential energy as a function 
of the position $\mathbf{q}'(t) 
= (\mathbf{q}'_{1}(t),\mathbf{q}'_{2}(t),\cdots,\mathbf{q}'_{N}(t))$,  
%(\equiv (q'_{jx}(t),q'_{jy}(t))), j=1,2,\cdots,N$, 
and the $\cald \times \cald$ matrix $\Theta\equiv(\Theta_{jk})$ 
is defined by 

\begin{eqnarray}
   \Theta_{jk} \equiv 
   \left\{
   \begin{array}{ll}
      1   & \mbox{in $j=1$ and $k=2$}  \\
      0   & \mbox{otherwise} 
   \end{array}
   \right. 
\end{eqnarray}

\noindent for $j=1,2,\cdots,\cald$ and $k=1,2,\cdots,\cald$.

   The second term on the right-hand side of 
Eq. (\ref{SllodThermEquat1}) is added so that
the quantity $\mathbf{p}'\equiv(\mathbf{p}'_{1}, 
\mathbf{p}'_{2},\cdots,\mathbf{p}'_{N})$ 
means the mass times the velocity in the 
moving frame by the velocity of the global shear flow, 
the so called  thermal momentum. 
   The second term on the right-hand side of 
Eq. (\ref{SllodThermEquat2}) is added so that we derive  
the equation 
%
%\begin{eqnarray}
$
   m d^{2}\mathbf{q}'_{j}(t)/dt^{2} = 
   - \partial U(\mathbf{q}'(t))/
   \partial\mathbf{q}'_{j}(t)
$
%   - \alpha(t) I  \mathbf{p}'_{j}(t)
%\label{SllodThermEquat1B}\end{eqnarray} 
%
%\noindent 
of the position $\mathbf{q}'(t)$ from Eqs. 
(\ref{SllodThermEquat1}), (\ref{SllodThermEquat2}) and 
$\Theta^{2}=0$.
   Therefore the parameter $\gamma$ dependence does not appear 
in the equation of $\mathbf{q}'(t)$ only. 
%only through the friction term $\alpha(t)$. 

   Usually the Sllod equation is used with a thermostat, 
such as the isokinetic thermostat \cite{Eva90a}, 
in order to make a model for the system driven 
by a shear rate with an attached heat 
reservoir which removes (as the house-keeping heat) the energy 
generated inside the system by the shear 
and maintains the temperature of the system constant in time. 
   The isokinetic thermostat is expressed by 
the term $- \alpha(t) \mathbf{p}'_{j}(t)$, which 
is added in the right-hand side of Eq. (\ref{SllodThermEquat2}). 
   Here $\alpha(t)$ is defined by 

\begin{eqnarray}
   \alpha(t) \equiv - 
   \frac{\sum_{j=1}^{N}\mathbf{p}'_{j}(t)^{T} 
   \left(\frac{\partial U(\mathbf{q}'(t))}
   {\partial\mathbf{q}'_{j}(t)} 
   + \gamma \Theta \mathbf{p}'_{j}(t)\right)}
   {\sum_{j=1}^{N} |\mathbf{p}'_{j}(t)|^{2}}
\label{ShearRate}\end{eqnarray}

\noindent as a function of $\mathbf{q}'_{j}(t)$ 
and $\mathbf{p}'_{j}(t) 
%(\equiv  (p'_{jx}(t),p'_{jy}(t)))
$ 
so that the total kinetic energy is constant in time: 
$d\bigl[\sum_{j=1}^{N}|\mathbf{p}'_{j}(t)|^{2}$ 
$/(2m)\bigr]/dt = 0$. 
   Note that adding the thermostated term 
$- \alpha(t) \mathbf{p}'_{j}(t)$ we ignore Galilean 
invariance of the dynamics. 
   Therefore the Sllod dynamics with the isokinetic 
thermostat is usually formulated in the center of mass 
frame and $\sum_{j=1}^{N} \mathbf{p}'_{j}(t) = 0$.
%   Equations (\ref{SllodThermEquat1}) and (\ref{SllodThermEquat2}) 
%are called the Sllod equation for the planar Coutte flow. 
%with the isokinetic thermostat expressed by 
%the term including the coefficient $\alpha(t)$. 
%and $I$ is the $\cald\times\cald$ identical matrix. 

   Now, we introduce 
$q'_{jx}$ ($q'_{jy}$) as the 
1st component (2nd component) of the coordinate variable 
$\mathbf{q}'_{j}$ of the $j$-th particle, and   
$p'_{jx}$ ($p'_{jy}$) as the 
1st component (2nd component) of the variable 
$\mathbf{p}'_{j}$ of the $j$-th particle.
   We define the operator $\hat{L}'$ by   

\begin{eqnarray}
   && i \hat{L}' X'(\bfGamma') 
   \nonumber \\
   &&\eqspaA = \sum_{j=1}^{N} 
   \left[ 
      \frac{\partial X'(\bfGamma')}{\partial\mathbf{q}'_{j}} 
      \cdot \frac{\partial H'(\bfGamma')}{\partial\mathbf{p}'_{j}} 
      - 
      \frac{\partial X'(\bfGamma')}{\partial \mathbf{p}'_{j}} 
      \cdot \frac{\partial H'(\bfGamma')}{\partial \mathbf{q}'_{j}} 
   \right]
   \nonumber \\
   &&\eqspaA \eqspaB
   + \gamma \sum_{j=1}^{N} 
   \left[
      \frac{\partial X'(\bfGamma')}{\partial q'_{jx}} q'_{jy}
      - 
      \frac{\partial X'(\bfGamma')}{\partial p'_{jx}} p'_{jy}
   \right] 
\label{SllodLiouv}\end{eqnarray}

\noindent for any function $X'(\bfGamma')$ of $\bfGamma'(\equiv 
(\mathbf{p}',\mathbf{q}'))$. 
   Here $H'(\bfGamma')$ is introduced as 

\begin{eqnarray}
%$
   H'(\bfGamma') \equiv 
   \sum_{j=1}^{N} 
      \frac{| \mathbf{p}'|^{2}}{2m}
      +U(\mathbf{q}') .
%$.
\label{SllodHamil}\end{eqnarray} 

\noindent As an important feature of the operator $i \hat{L}'$ 
we have 

\begin{eqnarray}
   i \hat{L}' H'(\bfGamma') 
   = -\gamma \mathcal{V} P_{xy} (\bfGamma')
\label{SllodLiouvHamil}\end{eqnarray}

\noindent using the function $P_{xy}(\bfGamma)$  
defined by Eq. (\ref{StresTenso0}). 
   Using the operator $i \hat{L}'$ defined by Eq. (\ref{SllodLiouv}),  
the Sllod dynamics (\ref{SllodThermEquat1}) and (\ref{SllodThermEquat2}) 
is simply represented as 
%
%\begin{eqnarray}
$
   d\bfGamma'(t)/dt = i \hat{L}'\bfGamma'(t)
$
%\label{SllodThermEquat3}\end{eqnarray} 
%
%\noindent 
for $\bfGamma'(t) \equiv 
\exp\{i \hat{L}'(t-t_{0})\}\bfGamma'=
(\mathbf{p}'(t),\mathbf{q}'(t))$, 
which is equivalent to the equation 

\begin{eqnarray}
   \frac{d g'(\bfGamma',t)}{dt} = - i \hat{L}'g'(\bfGamma',t)
\label{SllodThermEquat4}\end{eqnarray} 

\noindent for the distribution $g'(\bfGamma',t)$ of 
$\bfGamma'$ at time $t$, because of 
the relation $\int d\bfGamma' X(\bfGamma') g'(\bfGamma',t) 
= \int d\bfGamma' X(\bfGamma'(t)) $ $g'(\bfGamma',t_{0})$ 
for any function $X(\bfGamma')$ of $\bfGamma'$. 
   In other words, Eq. (\ref{SllodThermEquat4}) is 
the Liouville equation corresponding to the Sllod equation 
(\ref{SllodThermEquat1}) and (\ref{SllodThermEquat2}). 

   We define the distribution functions 
$g'_{0}(\bfGamma)$ and $\tilde{g}'(\bfGamma,t)$ as 
   
\begin{eqnarray}
   g'_{0}(\bfGamma')
   \equiv \Xi'{}^{-1} \exp\left\{
   -\beta H'(\bfGamma') 
   \right\}
\label{SllodShearCanon}\end{eqnarray}
%\begin{widetext}
\begin{eqnarray}
   \tilde{g}'(\bfGamma',t) 
      &\equiv& \exp \left\{-i\hat{L}' (t-t_{0})
      \right\} g'_{0}(\bfGamma')
      \label{SllodGenerShearCanon1} \\
   &=& \Xi' \exp\left\{
      -\beta \left[H'(\bfGamma') 
      + \int_{t_{0}}^{t} ds \right.\right.
      \nonumber \\
   && \left.\left. \times
      \frac{\partial 
      \exp \left\{-i\hat{L}' (s-t_{0})
      \right\} H'(\bfGamma')
      }{\partial s}
      \right]\right\} \nonumber \\
   &=& g'_{0}(\bfGamma') \exp\left\{ -\beta
      \gamma \mathcal{V} 
      \int_{t_{0}}^{t} ds \tilde{P}'_{xy}(\bfGamma',-s+2t_{0})
   \right\} .
      \nonumber \\
   && 
\label{SllodGenerShearCanon2}\end{eqnarray}
%\end{widetext}

\noindent where $\tilde{P}'_{xy}(\bfGamma',t)$ 
is defined by by $\tilde{P}'_{xy}(\bfGamma',t) 
\equiv\exp\{i\hat{L}'(t-t_{0})\} P_{xy}(\bfGamma')$ 
and $\Xi'$ is a normalization constant 
$\Xi'\equiv \int d\bfGamma' \exp\{
-\beta H'(\bfGamma') \}$. 
   Here we used the relation (\ref{SllodLiouvHamil}) 
to derive Eq. (\ref{SllodGenerShearCanon2}). 
   Using the distribution $\tilde{f}(\bfGamma,t)$ 
we define the average 
$\langle X(\bfGamma') \rangle_{t}'$ by 

\begin{eqnarray}
   \langle X(\bfGamma') \rangle_{t}' 
   \equiv 
   \int d\bfGamma X(\bfGamma') \tilde{g}'(\bfGamma',t) 
\label{ShearShearAvera}\end{eqnarray}

\noindent for any function $X(\bfGamma')$ of $\bfGamma'$.

   Now we discuss a relation between 
the above Sllod dynamics approach 
and the Hamiltonian dynamics approach 
of the text of this paper.   
   First, in the introduction of the Sllod equation 
we notice correspondences of the spatial coordinate and 
the velocity as 

\begin{eqnarray}
   \mathbf{p}' &\longleftrightarrow&  m\mathbf{v}^{(mov)} 
   \label{ShearCorres1a} \\
   \mathbf{q}' &\longleftrightarrow&  \mathbf{q}
\label{ShearCorres1b}\end{eqnarray}

\noindent where the quantities on the left-hand side 
are for the Sllod dynamics approach, and the ones 
on the right-hand side are for the Hamiltonian approach of the text. 
   Second, it is clear that Eqs. (\ref{SllodShearCanon}) and
(\ref{SllodGenerShearCanon2}) 
correspond to 
Eqs. (\ref{ShearCanon}) and (\ref{GenerShearCanon2}), 
respectively, therefore we also have correspondences as 

\begin{eqnarray}
   g'_{0}(\bfGamma') &\longleftrightarrow& f(\bfGamma) 
   \label{ShearCorres2a} \\
   i\hat{L}' &\longleftrightarrow& i\hat{L}^{(ine)} 
   \label{ShearCorres2b} \\
   \tilde{g}'(\bfGamma',t) 
      &\longleftrightarrow&  \tilde{f}(\bfGamma,t)   
   \label{ShearCorres2c} \\
   \langle \cdots \rangle_{t}' 
      &\longleftrightarrow& \langle \cdots \rangle_{t}
\label{ShearCorres2d}\end{eqnarray}
   
\noindent These correspondences are not mathematical 
equivalences, but we can discuss some physical meanings 
in them. 
   As an example, let's consider the correspondence 
(\ref{ShearCorres2a}) more concretely. 
   We introduce the new variables 
$\tilde{\mathbf{p}}'_{j} \equiv 
\mathbf{p}'_{j} + \gamma \Theta \mathbf{q}'_{j}$, 
and $\tilde{\bfGamma}'\equiv 
(\tilde{\mathbf{p}}',\mathbf{q}')$ with 
$\tilde{\mathbf{p}}' 
\equiv (\tilde{\mathbf{p}}'_{1},\tilde{\mathbf{p}}'_{2},
\cdots,\tilde{\mathbf{p}}'_{N})$. 
   Here, $\tilde{\bfGamma}'$ is the vector for the Sllod equation, 
 which corresponds to $\bfGamma$ for the approach of the text: 
$\tilde{\bfGamma}' \longleftrightarrow \bfGamma$. 
   Using this vector $\tilde{\bfGamma}'$ we 
rewrite the distribution $g'_{0}(\bfGamma')$ as

\begin{eqnarray}
   && 
   g'_{0}(\bfGamma')
      \nonumber \\
   &&  
   = \Xi'{}^{-1} \exp\left\{
   -\beta \left[
     H^{(mov)}(\tilde{\bfGamma}') 
     + \sum_{j=1}^{N} \frac{1}{2} m 
     (\gamma q'_{jy})^{2}
   \right]
   \right\}
      \nonumber \\
\label{SllodShearCanon2}\end{eqnarray}

\noindent where we used the function form 
$H^{(mov)}(\bfGamma)$ of $\bfGamma$ 
given by Eq. (\ref{ShearHamilMov}). 
   The Sllod dynamics approach does not take into 
account the inertial force, 
as same as the approach 
using the distribution (\ref{CanonLocalEquib}), 
%under the local equilibrium assumption, 
and this is the reason why the second term 
$\sum_{j=1}^{N} m 
(\gamma q'_{jy})^{2}/2$
 in the square bracket of the right-hand side of 
 Eq. (\ref{SllodShearCanon2}), 
which make a difference of $g'_{0}(\bfGamma')$ 
from $f(\bfGamma) = \Xi^{-1} \exp[
-\beta H^{(mov)}(\bfGamma) ]$, appears. 
   However as far as we can neglect the effect of 
the inertial force, 
%by comparing Eq. (\ref{SllodShearCanon2}) with 
%Eq. (\ref{Canon0}) 
the distribution $g'_{0}(\bfGamma')$ 
can correspond to the distribution $f(\bfGamma)$ 
in the shear flow system, namely we obtain the 
correspondence (\ref{ShearCorres2a}). 

   It should be noted that the vector $\bfGamma'$ used in 
the Sllod equation does not correspond 
to the phase space vector $\bfGamma$, so the quantity 
$P_{xy} (\bfGamma')$ on the right-hand side of Eq. 
(\ref{SllodLiouvHamil}) does not generally correspond to 
the quantity $P_{xy} (\bfGamma)$ on the right-hand side 
of Eq. (\ref{HelfaTimDer}). 
   Actually, noting the corresponding (\ref{ShearCorres1a}) 
we notice 

\begin{eqnarray}
   P_{xy} (\bfGamma') 
      + \frac{\gamma}{\mathcal{V}}\sum_{j=1}^{N} q'_{jy} p'_{jy} 
   &\longleftrightarrow&  
   P_{xy} (\bfGamma) .
\label{ShearCorres3}\end{eqnarray}

\noindent 
   However, if the two quantities $q'_{jy}$ and $p'_{jy}$ are 
decoupled, namely $\langle q'_{jy}p'_{jy} \rangle_{\infty}' 
\approx \langle q'_{jy} \rangle_{\infty}' 
\langle p'_{jy} \rangle_{\infty}'$,  
and the average $\langle p'_{jy} \rangle_{\infty}'$ is zero, 
then we can have an approximate correspondence 
$\langle P_{xy} (\bfGamma') \rangle_{\infty}' 
\longleftrightarrow \langle P_{xy} (\bfGamma)\rangle_{\infty}$. 
%   As another important relation between the Sllod equation 
%approach and the approach used in the text, 

   We can also show that using the distribution 
$\tilde{g}'(\bfGamma',t)$ 
based on the Sllod equation, we can calculate the viscosity 
$\eta$ as 

\begin{eqnarray}
   \eta &=& - \lim_{\gamma\rightarrow 0} 
      \frac{\left\langle P_{xy}(\bfGamma') \right 
      \rangle_{\infty}'}{\gamma}
      \nonumber \\
   &=& -\lim_{\gamma\rightarrow 0} 
      \frac{\int d \bfGamma' P_{xy}(\bfGamma') g'_{0}(\bfGamma')}{\gamma}
      \nonumber \\
   && \eqspaB
      + \lim_{\gamma\rightarrow 0} \beta
      \mathcal{V}
      \int_{t_{0}}^{\infty} ds \int d\bfGamma' 
      P_{xy}(\bfGamma')  
      \nonumber \\
   && \eqspaA \eqspaB \times \tilde{P}'_{xy}(\bfGamma',-s+2t_{0}) 
      g'_{0}(\bfGamma')
      \nonumber \\
   &=& \beta \mathcal{V} 
      \int_{t_{0}}^{\infty} dt \;\; 
      \left\langle \tilde{P}_{xy}(\bfGamma,t) 
      P_{xy}(\bfGamma) 
      \right\rangle^{(eq)} .
\label{ShearGreenKuboVisco}\end{eqnarray}

\noindent where we used 

\begin{eqnarray}
   \int d \bfGamma' P_{xy}(\bfGamma') g'_{0}(\bfGamma')
      = 0
\label{Sllod0thviscoDer1}\end{eqnarray}

\noindent whose derivation is similar to that 
of Eq. (\ref{zerothvisco}), 
noting the correspondences 
(\ref{ShearCorres2a}) and (\ref{ShearCorres2b}).
   Therefore, using the Sllod dynamics approach 
we obtain the same linear response 
formula for viscosity as Eq. (\ref{GreenKuboVisco}).

%%%%%%%%%%%%%%%%%%%%%%%%%%%%%%%%%%%%%%%%%%%%%%%%%%%%%%%%%%%%%%%%%%%%%%

\section{Second Law of Thermodynamics in the Non-Equilibrium Canonical Distribution approach}
\label{AppenSecondLaw}

   In this appendix we give a derivation of the inequality 
(\ref{SecondLaw}) satisfied at any time $t \; (>t_{0})$. 

   We start our derivation from the inequality
   
\begin{eqnarray}
   x \ln x -x +1 \geq 0 
\label{logx}\end{eqnarray}

\noindent satisfied by any positive real number $x \; (>0)$. 
   The equality in (\ref{logx}) is satisfied only when $x=1$. 
   Using the inequality (\ref{logx}) in the case 
$x=\tilde{f}(\bfGamma,t)/f(\bfGamma)$ we have 
   
\begin{eqnarray}
   \frac{\tilde{f}(\bfGamma,t)}{f(\bfGamma)} 
   \ln \frac{\tilde{f}(\bfGamma,t)}{f(\bfGamma)} 
   - \frac{\tilde{f}(\bfGamma,t)}{f(\bfGamma)}  
   + 1   \geq 0 ,  
\label{logx2}\end{eqnarray}

\noindent which is equivalent to 
   
\begin{eqnarray}
   \tilde{f}(\bfGamma,t) \ln \tilde{f}(\bfGamma,t) 
   - \tilde{f}(\bfGamma,t) \ln f(\bfGamma) 
   \geq \tilde{f}(\bfGamma,t) - f(\bfGamma) .  
   \nonumber \\
\label{logx2b}\end{eqnarray}

\noindent Now we note  

\begin{eqnarray}
   \int d\bfGamma \tilde{f}(\bfGamma,t) 
   = \int d\bfGamma f(\bfGamma) \; (= 1),   
\label{integff}\end{eqnarray}
\begin{eqnarray}
   && \int d\bfGamma 
      \tilde{f}(\bfGamma,t) \ln \tilde{f}(\bfGamma,t) 
      \nonumber \\
   && \eqspaA = \int d\bfGamma e^{-i\hat{L}^{(ine)}(t-t_{0})} 
      \left[f(\bfGamma) \ln f(\bfGamma) \right]  
      \nonumber \\
   && \eqspaA  = \int d\bfGamma 
      f(\bfGamma) \ln f(\bfGamma).  
\label{xlonxtime}\end{eqnarray}

\noindent By taking integral with respect to $\bfGamma$ 
on both sides of the inequality (\ref{logx2b}), 
and by using Eqs. (\ref{ObserEntro}), 
(\ref{integff}) and (\ref{xlonxtime}) we obtain 

\begin{eqnarray}
   \int d\bfGamma\tilde{f}(\bfGamma,t) S(\bfGamma)
   \geq  \int d\bfGamma 
      f(\bfGamma) S(\bfGamma) .  
\label{logx3}\end{eqnarray}

\noindent Using the equation 
%Eqs. (\ref{GenEntro}) and 
$\overline{S} = \int d\bfGamma 
f(\bfGamma) S(\bfGamma) = \langle S \rangle_{t_{0}}$
in the inequality (\ref{logx3}), 
we obtain the inequality (\ref{SecondLaw}).

%%%%%%%%%%%%%%%%%%%%%%%%%%%%%%%%%%%%%%%%%%%%%%%%%%%%%%%%%%%%%%%%%%%%%%

\section{Stability Condition for the Shear Flow}
\label{ShearStabiCondi}

   In this appendix we show the equivalence between the 
condition (\ref{ShearStabi2}) and the conditions
	(\ref{ShearStabi3a}) and (\ref{ShearStabi3b}). 
   We also give a derivation of Eq. (\ref{ShearStabi3c}). 

   Noting that the energy 
$\overline{H^{(ine)}}$ is the function of $\overline{S}$ 
and $\overline{Q}$ by Eq. (\ref{SearFirstLaw4}), we have 

\begin{eqnarray}
   \delta T = \delta \frac{\partial \overline{H^{(ine)}}}
      {\partial \overline{S}} 
%   \nonumber \\
%   &=& 
   = \frac{\partial^{2} \overline{H^{(ine)}}}
      {\partial \overline{S}^{2}} \delta \overline{S}
      + \frac{\partial^{2} \overline{H^{(ine)}}}
      {\partial \overline{S}\partial \overline{Q}} 
      \delta \overline{Q} 
%      \nonumber \\
\label{AppDeriv1}\end{eqnarray} 

\noindent Using Eq. (\ref{AppDeriv1}) we also have 

%\begin{widetext}
\begin{eqnarray}
   \delta \gamma &=& \delta \frac{\partial \overline{H^{(ine)}}}
      {\partial \overline{Q}} \nonumber \\
   &=& \frac{\partial^{2} \overline{H^{(ine)}}}
      {\partial \overline{S}\partial \overline{Q}} 
      \delta \overline{S}
      + \frac{\partial^{2} \overline{H^{(ine)}}}
      {\partial \overline{Q}^{2}} \delta \overline{Q} \nonumber \\
   &=& \frac{\partial^{2} \overline{H^{(ine)}}}
      {\partial \overline{S}\partial \overline{Q}} 
      \left(\frac{\partial^{2} \overline{H^{(ine)}}}
      {\partial \overline{S}^{2}} \right)^{-1}       
      \left[\delta T- \frac{\partial^{2} \overline{H^{(ine)}}}
      {\partial \overline{S}\partial \overline{Q}} 
      \delta \overline{Q}\right]
      \nonumber \\
   &&\eqspaA\eqspaB + \frac{\partial^{2} \overline{H^{(ine)}}}
      {\partial \overline{Q}^{2}} \delta \overline{Q}
      \nonumber \\
   &=& \left(\frac{\partial^{2} \overline{H^{(ine)}}}
      {\partial \overline{S}^{2}} \right)^{-1} 
      \frac{\partial^{2} \overline{H^{(ine)}}}
      {\partial \overline{S}\partial \overline{Q}} 
      \delta T
      \nonumber \\
   &&\hspace{0cm}   
      +\left[ 
      \frac{\partial^{2} \overline{H^{(ine)}}}
      {\partial \overline{Q}^{2}}-\left(\frac{\partial^{2} 
      \overline{H^{(ine)}}}
      {\partial \overline{S}^{2}} \right)^{-1} 
      \left(\frac{\partial^{2} \overline{H^{(ine)}}}
      {\partial \overline{S}\partial \overline{Q}} 
      \right)^{2}
      \right] \delta \overline{Q} , 
      \nonumber \\
\label{AppDeriv2}\end{eqnarray} 

\noindent which leads to 

\begin{eqnarray}
   \left.\frac{\partial\gamma}{\partial \overline{Q}}\right|_{T} 
    =
      \frac{\partial^{2} \overline{H^{(ine)}}}
      {\partial \overline{Q}^{2}}-\left(\frac{\partial^{2} 
      \overline{H^{(ine)}}}
      {\partial \overline{S}^{2}} \right)^{-1} 
      \left(\frac{\partial^{2} \overline{H^{(ine)}}}
      {\partial \overline{S}\partial \overline{Q}} 
      \right)^{2} .
      \nonumber \\
\label{AppDeriv3}\end{eqnarray} 

\noindent Using Eq. (\ref{AppDeriv1}) and  (\ref{AppDeriv3}) 
we obtain

\begin{eqnarray}
   \delta^{2} \overline{H^{(ine)}} 
   &=& \frac{\partial^{2} \overline{H^{(ine)}}}
      {\partial \overline{S}^{2}} (\delta \overline{S})^{2}
   + 2 \frac{\partial^{2} \overline{H^{(ine)}}}
      {\partial \overline{S}\partial \overline{Q}}
      \delta \overline{S} \delta \overline{Q}
      \nonumber \\
   &&\eqspaA\eqspaB
   + \frac{\partial^{2} \overline{H^{(ine)}}}
      {\partial \overline{Q}^{2}} (\delta \overline{Q})^{2} 
      \nonumber \\
   &=& \left( \delta T
   + \frac{\partial^{2} \overline{H^{(ine)}}}
      {\partial \overline{S}\partial \overline{Q}} \delta \overline{Q}
      \right) \delta \overline{S}
%      \nonumber \\
%   &&\eqspaA\eqspaB
   + \frac{\partial^{2} \overline{H^{(ine)}}}
      {\partial \overline{Q}^{2}} (\delta \overline{Q})^{2} 
      \nonumber \\
   &=&\left(\frac{\partial^{2} \overline{H^{(ine)}}}
      {\partial \overline{S}^{2}}\right)^{-1}
      (\delta T)^{2}
   + \left[\frac{\partial^{2} \overline{H^{(ine)}}}
      {\partial \overline{Q}^{2}} \right.
      \nonumber \\
   &&\eqspaA \left.
      - \left(\frac{\partial^{2} \overline{H^{(ine)}}}
      {\partial \overline{S}^{2}}\right)^{-1}
      \left(\frac{\partial^{2} \overline{H^{(ine)}}}
      {\partial \overline{S}\partial \overline{Q}}
      \right)^{2}
      \right]
      (\delta \overline{Q})^{2} 
      \nonumber \\
   &=&\left(\left.\frac{\partial T}
      {\partial \overline{S}}\right|_{\overline{Q}} \right)^{-1}
      (\delta T)^{2} 
      + \left.\frac{\partial\gamma}{\partial \overline{Q}}\right|_{T} 
      (\delta\overline{Q})^{2} .
\label{ShearStabi3}\end{eqnarray} 

\noindent The inequality (\ref{ShearStabi2}) must be satisfied 
by any infinitesimal deviations $\delta T$ and $\delta \overline{Q}$, 
so using Eq. (\ref{ShearStabi3}) 
we obtain the conditions  (\ref{ShearStabi3a}) 
and (\ref{ShearStabi3b}). 

   Now, using the canonical distribution 
(\ref{ShearCanon}), we calculate the derivative of $\overline{Q}$ 
with respect to $\gamma$ at constant temperature $T$.   

\begin{eqnarray}
   \left.\frac{\partial\overline{Q}}{\partial \gamma}\right|_{T} 
   &=& \frac{ \partial }{\partial \gamma} 
   \Xi^{-1} \int d\bfGamma  Q(\bfGamma) 
      \nonumber \\
   &&\eqspaA\eqspaB \times
     \exp\left\{-\beta \left[
      H^{(ine)}(\bfGamma) - \gamma Q(\bfGamma) 
      \right]\right\} \nonumber \\ 
   &=& \beta \Xi^{-1} \int d\bfGamma  \left[Q(\bfGamma)\right]^{2} 
      \nonumber \\
   &&\eqspaA\eqspaB \times
      \exp\left\{-\beta \left[
      H^{(ine)}(\bfGamma) - \gamma Q(\bfGamma) 
      \right]\right\} 
   \nonumber \\ 
   &&\eqspaA 
   - \Xi^{-2} \frac{\partial \Xi}{\partial \gamma} 
      \int d\bfGamma  Q(\bfGamma) 
      \nonumber \\
   &&\eqspaA\eqspaB \times
      \exp\left\{-\beta \left[
      H^{(ine)}(\bfGamma) - \gamma Q(\bfGamma) 
      \right]\right\} \nonumber \\
   &=& \beta \left(\overline{Q^{2}}- \overline{Q}^{2}\right)
\label{AppShearStabi3c1}\end{eqnarray} 
%\end{widetext}

\noindent where we used 
%
%\begin{eqnarray}
$
   \Xi^{-1} \partial \Xi/\partial \gamma
   = \beta \overline{Q}
$.
%\label{AppShearStabi3c2}\end{eqnarray} 
%
Therefore we obtain Eq. (\ref{ShearStabi3c}). 

%%%%%%%%%%%%%%%%%%%%%%%%%%%%%%%%%%%%%%%%%%%%%%%%%%%%%%%%%%%%%%%%%%%%%%

\section{Relation Between the Two Averages}
\label{TwoAveRel}

   In this appendix we give a derivation of Eq. (\ref{TwoAvera}). 
   
Using the expression (\ref{GenerShearCanon2}) for the 
distribution $\tilde{f}(\bfGamma,t)$ used 
in the average $\langle X(\bfGamma) \rangle_{t}$ 
for any function $X(\bfGamma)$, 
we have

\begin{eqnarray}
   &&\langle X \rangle_{t} 
      \nonumber \\
   &&= \overline{X} + \left\langle \left(
      X - \overline{X}\right)\right\rangle_{t}
      \nonumber \\
   &&= \overline{X} + \int_{t_{0}}^{t} ds 
      \frac{\partial \left\langle \left(
      X - \overline{X}\right)\right\rangle_{s}}{\partial s}
      \nonumber \\
   &&= \overline{X} 
      - \beta \gamma \mathcal{V} \int_{t_{0}}^{t} ds 
      \left\langle 
         \left[X(\bfGamma) - \overline{X}\right]
         \tilde{P}_{xy}(\bfGamma,-s+2t_{0}) 
      \right\rangle_{s} 
      \nonumber \\
   &&= \overline{X} 
      - \beta \gamma \mathcal{V} \int_{t_{0}}^{t} ds 
      \overline{\left[
         \tilde{X}(\bfGamma,s) - \overline{X}
      \right]P_{xy}(\bfGamma)}
      \nonumber \\
   &&= \overline{X} 
      - \beta \gamma \mathcal{V} \int_{t_{0}}^{t} ds 
      \overline{\left[
         \tilde{X}(\bfGamma,s) - \overline{X}
      \right]
      \Bigl[P_{xy}(\bfGamma) -  \overline{P_{xy}} \; \Bigr]
      }, 
      \nonumber \\
   &&
\label{TwoAvera2}\end{eqnarray} 

\noindent where we used Eqs. (\ref{zerothvisco}),  
$\langle (X - \overline{X})\rangle_{t_{0}} = 0$ 
and $(i \hat{L}^{(ine)})^{\dagger} = -i\hat{L}^{(ine)}$. 
   By taking the limit $t\rightarrow\infty$ 
in Eq. (\ref{TwoAvera2}), we obtain Eq. (\ref{TwoAvera}). 
   Concerning Eq. (\ref{TwoAvera2}) one may notice 

\begin{eqnarray}
   \overline{\left[
         \tilde{X}(\bfGamma,s) - \overline{X}
      \right]
      \Bigl[P_{xy}(\bfGamma) -  \overline{P_{xy}} \; \Bigr]
      }
      =
   \overline{
         \tilde{X}(\bfGamma,s) 
      P_{xy}(\bfGamma) 
      }
      \nonumber \\
\label{TwoAvera3}\end{eqnarray} 

\noindent because of Eq. (\ref{zerothvisco}),
so the integral function in the second tem of the 
right-hand side of Eq. (\ref{TwoAvera2}) can be replaced by the 
right-hand side of Eq. (\ref{TwoAvera3}).

%%%%%%%%%%%%%%%%%%%%%%%%%%%%%%%%%%%%%%%%%%%%%%%%%%%%%%%%%%%%%%%%%%%%%%

%%%%%%%%%%%%%%%%%%%%%%%%%%%%%%%%%%%%%%%%%%%%%%%%%%%%%%%%%%%%%%%%%%%%%%


\begin{thebibliography}{99}
\bibitem{noteI1}  Evans and Hanley also added the term 
$-\mathcal{P} d\mathcal{V}$ on the right-hand side 
of Eq. (\ref{EvansHanley}) 
to discuss the pressure $\mathcal{P}$ as an energy response to a 
change of the volume $\mathcal{V}$ in shear flow, 
by analogy with equilibrium thermodynamics. 
   However it is not evident that we can justify 
such a term in non-equilibrium steady states at the level of the 
theory presented here. 
   We mention some problems with the discussion of the pressure 
briefly in Sec. \ref{ConclusionRemarks}.
   In this paper we mainly concentrate on the  
fixed volume case where $d\mathcal{V} = 0$.
\bibitem{note2} A non-equilibrium molecular algorithm 
%for shear flow 
derived from a Hamiltonian,  
like the one derived from the Hamiltonian 
$H^{(mov)}(\bfGamma)$ of Eq. (\ref{ShearHamilMov}), 
is the so-called Dolls tensor algorithm \cite{Eva90a}. 
   However we should notice some differences 
between the dynamics derived from 
$H^{(mov)}(\bfGamma)$ and from the Dolls tensor Hamiltonian. 
   For example, the momentum used in 
the dynamics derived from 
$H^{(mov)}(\bfGamma)$ is the phase space momentum 
justified by the Lagrangian formalism, 
while the momentum used in the Dolls tensor dynamics is 
the thermal momentum, namely the mass times the velocity 
in the moving frame (defined by the global velocity profile).
\bibitem{noteIV1} The divergence of a non-equilibrium entropy in a steady 
shear flow in thermostated dynamics, has been considered
for example see Ref. \cite{Eva89}. 
   However, note that the explanation for the entropy divergence in 
that case is different from that used here.
\bibitem{note3} As a related point, one 
can notice that the canonical distribution $f(\bfGamma)$ 
can be derived from a local minimum condition 
for the entropy $\overline{S}= - \int d\bfGamma f(\bfGamma) 
\ln f(\bfGamma)$ under the constraints of 
$\overline{H^{(int)}} = const$, $\overline{Q} = const$ and 
$1 = const$. 
   Actually, using the method of Lagrange's undetermined 
multipliers  with constants $a_{j}$, $j=1,2,3$ we have 
$0=\delta (\overline{S}+a_{1}\overline{H^{(int)}} 
+a_{2}\overline{Q} +a_{3})$, 
namely $0=-\ln f(\bfGamma) + 
1 + a_{1}H^{(int)}(\bfGamma)+a_{2}Q(\bfGamma) +a_{3}$. 
   In the case where $a_{1}=-\beta$, $a_{2}=\beta\gamma$ 
and $a_{3}=-\ln\Xi - 1$ we obtain the form (\ref{ShearCanon}) 
for the canonical distribution for shear flow.
\end{thebibliography}
\end{document}